\documentclass[amssymb,nobibnotes,superscriptaddress,longbibliography,twocolumn]{revtex4-1}

\usepackage{enumerate,amsmath}
\usepackage{graphicx}
\usepackage{color}

\begin{document}

\title{Harnessing disordered ensemble quantum dynamics for machine learning}%

\author{Keisuke Fujii}
\address{Photon Science Center, Graduate School of Engineering,
The University of Tokyo, 2-11-16 Yayoi, Bunkyo-ku, Tokyo 113-8656, Japan}
\address{The Hakubi Center for Advanced Research, Kyoto University, Yoshida-Ushinomiya-cho, Sakyo-ku, Kyoto 
606-8302, Japan}
\address{Department of Physics, Graduate School of Science,
Kyoto University, Kitashirakawa Oiwake-cho,
Sakyo-ku, Kyoto 606-8502, Japan}
\address{JST, PRESTO, 4-1-8 Honcho, Kawaguchi, Saitama 332-0012, Japan}

\author{Kohei Nakajima}
\address{The Hakubi Center for Advanced Research, Kyoto University, Yoshida-Ushinomiya-cho, Sakyo-ku, Kyoto 606-8302, Japan}
\address{JST, PRESTO, 4-1-8 Honcho, Kawaguchi, Saitama 332-0012, Japan}
\address{Graduate School of Informatics, Kyoto University, Yoshida Honmachi, Sakyo-ku, Kyoto 606-8501, Japan}

\date{\today}
\begin{abstract}
Quantum computer has an amazing potential of fast information processing. However, realisation of a digital quantum computer is still
a challenging problem
requiring highly accurate controls and key application strategies. Here we propose a novel platform, 
quantum reservoir computing, to solve these issues successfully by exploiting natural quantum dynamics
of ensemble systems, which is ubiquitous in laboratories nowadays, for machine learning. 
This framework enables ensemble quantum systems to universally emulate nonlinear dynamical systems including classical chaos.
A number of numerical experiments show that quantum systems consisting of 5--7 qubits possess computational capabilities comparable to conventional recurrent neural networks of 100--500 nodes. 
This discovery opens up a new paradigm for information 
processing with artificial intelligence powered by quantum physics.
\end{abstract}

\maketitle
\section{Introduction}

Quantum physics, which is the fundamental framework of physics, exhibits rich dynamics, sufficient to explain natural phenomena in microscopic worlds.
As Feynman pointed out~\cite{Feynman}, the simulation of quantum systems on classical computers is extremely challenging because of the high complexity of these systems.
Instead, they should be simulated by using a machine of which the operation is based on the laws of quantum physics.

Motivated by the recent rapid experimental progress in controlling complex quantum systems, non-conventional information processing utilising quantum physics has been explored in the field of quantum information science~\cite{NielsenChuang,FujiiText}.
For example,  certain mathematical problems, such as integer factorisation, which are believed to be intractable on a classical computer, are known to be efficiently solvable by a sophisticatedly synthesized quantum algorithm~\cite{ShorFact}.
Therefore, considerable experimental effort has been devoted to realising full-fledged universal quantum computers~\cite{UCSB1,UCSB2}.
On the other hand, quantum simulators are thought to be much easier to implement than a full-fledged universal quantum computer.
In this regard, existing quantum simulators have already shed new light on the physics of complex many-body quantum systems~\cite{QuantumSimulator1,QuantumSimulator2,QuantumSimulator3}, and a restricted class of quantum dynamics, known as adiabatic dynamics, has also been applied to combinatorial optimisation problems~\cite{Nishimori,Farhi,Dwave,Dwave2}.
However, complex real-time quantum dynamics, which is one of the most difficult tasks for classical computers to simulate
~\cite{FujiiMorimae,Fujiietal,FujiiTamate} 
and has great potential to perform nontrivial information processing, is now waiting to be harnessed as a resource for more general purpose information processing.
Specifically, the recent rapid progress in sensing and Internet technologies has resulted in an increasing demand for fast intelligent big data analysis with low energy consumption.
This has motivated us to develop brain-inspired information processing devices of a non-von Neumann type, on which machine learning tasks are able to run natively~\cite{IBM_Neuro}.

Here we propose a novel framework to exploit the complexity of real-time quantum dynamics
in ensemble quantum systems for nonlinear and temporal learning problems.
These problems include a variety of real-world tasks such as time-dependent signal processing, speech recognition, natural language processing, sequential motor control of robots, and stock market predictions.
Our approach is based on a machine learning technique inspired by the way the brain processes information, so-called {\it reservoir computing}~\cite{Jaeger0,Maass0,Reservoir}. 
In particular, this framework focuses on real-time computing with time-varying input that requires the use of memory, unlike feedforward neural networks.
In this framework, the low-dimensional input is projected to a high-dimensional dynamical system, which is typically referred to as a {\it reservoir}, generating transient dynamics that facilitates the separation of input states \cite{Transient}.
If the dynamics of the reservoir involve both adequate memory and nonlinearity \cite{Capacity}, emulating nonlinear dynamical systems only requires adding a linear and static readout from the high-dimensional state space of the reservoir.

A number of different implementations of reservoirs have been proposed, such as abstract dynamical systems for echo state networks (ESNs) \cite{Jaeger0} or models of neurons for liquid state machines \cite{Maass0}.
The implementations are not limited to programs running on the PC but also include physical systems, such as the surface of water in a laminar state \cite{Bucket}, analogue circuits and optoelectronic systems \cite{Laser0,Laser1,Laser1b,Laser2,Laser3,Laser4}, and neuromorphic chips \cite{Neuromorphic0}.
Recently, it has been reported that the mechanical bodies of soft and compliant robots have also been successfully used as a reservoir \cite{Helmut0,Kohei0,Kohei1,Kohei2,Kohei3,Ken1}.
In contrast to the refinements required by learning algorithms, such as in deep learning \cite{DL}, the approach followed by reservoir computing, especially when applied to real systems, is to find an appropriate form of physics that exhibits rich dynamics, thereby allowing us to outsource a part of the computation.
Nevertheless, no quantum physical system has been employed yet as a physical reservoir.

Here we formulate quantum reservoir computing (QRC) and show, through a number of numerical experiments, that disordered quantum dynamics can be used as a powerful reservoir.
Although there have been several prominent proposals on utilising quantum physics in the context of machine learning~\cite{Briegel12,QauntumDeep,Briegel14,Lloyed14,LloyedNatPhys,QLearning}, they are based on sophisticatedly synthesised quantum circuits on a full-fledged universal quantum computer.
Contrary to these software approaches, the approach followed by QRC is to exploit the complexity of natural (disordered) 
quantum dynamics for information processing, as it is.
Here disordered quantum dynamics means that
couplings are random, and hence 
no fine tuning of the parameters of the Hamiltonian
is required. Any quantum chaotic (non-integrable) system can be 
harnessed, and its computational capabilities are specified.
This is a great advantage, because we can utilise existing quantum simulators or complex quantum systems as resources to boost information processing.
Among existing works on quantum machine learning~\cite{Briegel12,Briegel14,Lloyed14,LloyedNatPhys,QLearning}, our approach is the first attempt to exploit quantum systems for temporal machine learning tasks, which essentially require a memory effect to the system.
As we will see below, our benchmark results show that quantum systems consisting of 5--7 qubits already exhibit a powerful performance comparable to the ESNs of 100-500 nodes.
Not only its computational power, 
QRC will also provide us an operational means
to approach complex real-time quantum dynamics.
While there had been a missing operational link between 
classical chaos and quantum chaotic systems
manifested by a Wigner-Dyson type statistics of the energy level spacing~\cite{Wigner,RMT},
it is quite natural to connect them
via the QRC framework naturally 
as an emulation of classical chaos by quantum chaotic systems.
Moreover, since complex quantum dynamics is ubiquitous,
this framework provides us new operational understanding of quantum physics,
such as quantum simulation, thermodynamics in closed quantum system and 
fast scrambling in black hole.

\section{Quantum reservoir computing}
\begin{figure}
\centering
\includegraphics[width=80mm,clip]{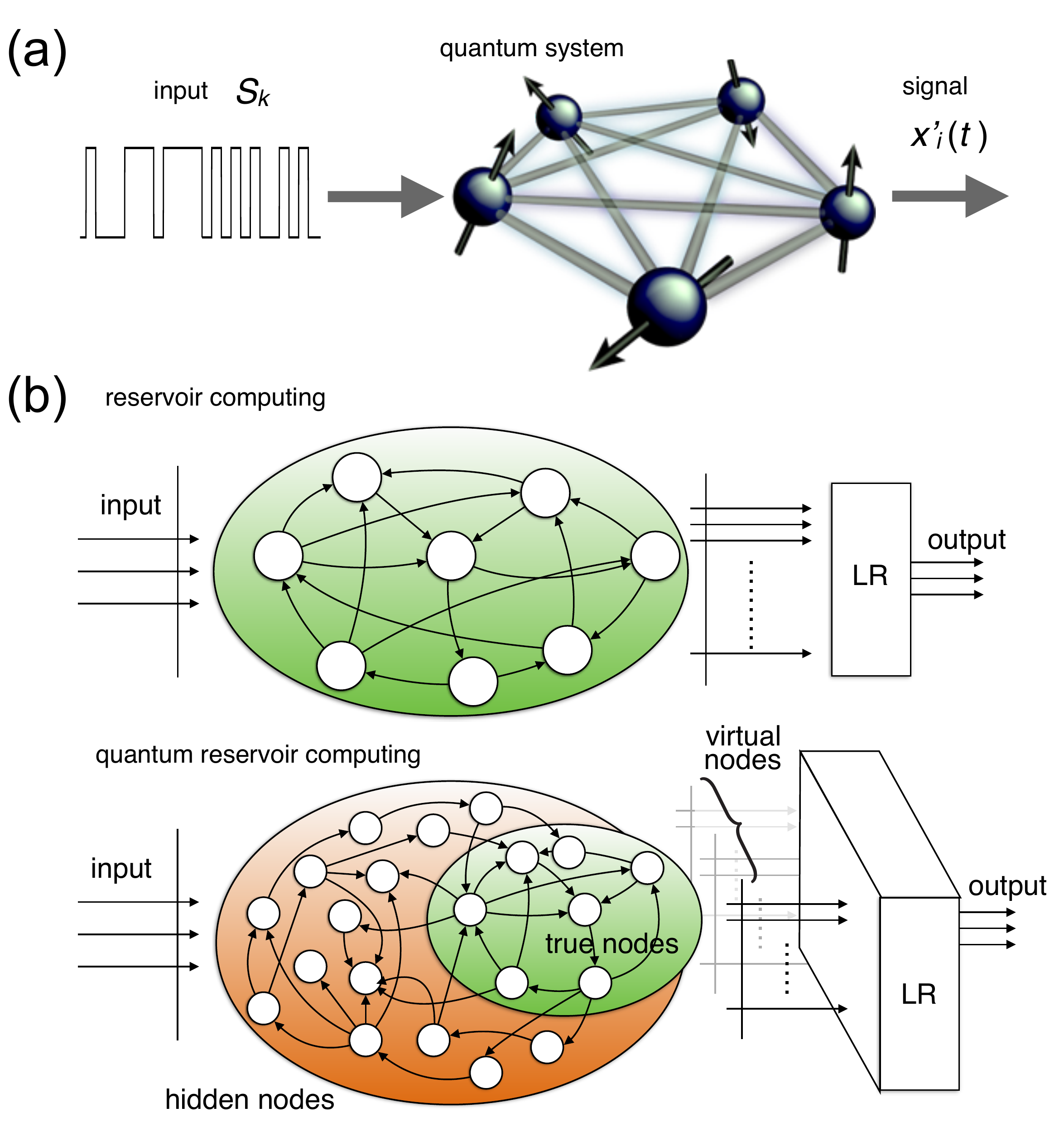}
\caption{
Information processing scheme in QRC. (a) The input sequence $\{s_k\}$ is injected into the quantum system.
The signal $x'_i (t)$ is obtained from each qubit.
(b) Comparison between conventional (upper) and quantum (lower) reservoir computing approaches.
Note that the circles in the QRC do not represent qubits,
but the basis of the Hilbert space like the nodes in quantum walk~\cite{Briegel14,QWalk1,QWalk2}.
The true nodes correspond to a subset of basis of the operator space
that are directly monitored by the ensemble measurements.
The hidden nodes correspond to the remaining degrees of freedom.
}
\label{figxx}
\end{figure}
\subsection{Description of quantum system and dynamics}
In this subsection, we will explain how to describe 
quantum system and dynamics for the readers who are not
familiar with quantum information.
The minimum unit of information in 
quantum physics is a quantum bit (qubit),
which consists of a two-level quantum system,
namely a vector 
in a two-dimensional complex vector space of spanned by $\{ |0\rangle, |1\rangle\}$.
Let us consider a quantum system consisting of $N$ qubits,
which is described as a tensor product space of a complex vector space of two dimensions.
A pure quantum state is represented by 
a state vector $|\psi \rangle$ in a $2^N$-dimensional 
complex vector space.
We may also consider a statistical (classical)
mixture of the states of the pure states,
which can be described by a $2^N \times 2^N$ hermitian matrix $\rho$ known as a density matrix.
For a closed quantum system, the time evolution for a time interval $\tau$ is 
given by a unitary operator $e^{-i H\tau}$
generated by a hermitian operator $H$ called Hamiltonian.
Specifically,
for the density matrix the time evolution is given by 
\begin{eqnarray}
\rho (t+ \tau) = e^{-i H \tau} \rho (t) e^{i H \tau},
\end{eqnarray}
where the Hamiltonian $H$ is an $2^N \times 2^N$ hermitian matrix 
and defines the dynamics of the quantum system.

\subsection{Measurements in ensemble quantum systems}
Measurements in quantum system is described by 
a set of projective operators $\{ P_i \}$,
which satisfies $\sum _i P_i = I$ and $P_{i} P_{j}=\delta _{ij} P_i$.
Then the probability to obtain the measurement outcome $i$
for the state $\rho$
is given by $p_i = {\rm Tr}[P_i \rho ]$.
The state after the measurement gets 
a backaction and is
given by $P_i \rho P_i /{\rm Tr[P_i \rho ]}$.
That is, a single quantum system inevitably disturbed 
by the projective measurement.
By repeating the projective measurements,
we can calculate average values $\langle O \rangle := {\rm Tr} [O\rho]$ 
of an observable $O = \sum _i a_i P_i $.

Here we consider an ensemble quantum system,
where the system consists of a huge number of the copies of $\rho$,
i.e., $\rho ^{\otimes m}$.
Cold atomic ensembles and liquid or solid state molecules
are natural candidates of such an ensemble quantum system.
For example, in an NMR (nuclear magnetic resonance) spin ensemble system,
we have typically $10^{18-20}$ copies of the same molecules~\cite{NMRQC1,NMRQC2}.
Nuclear spin degree of freedoms of them can be employed as the quantum system,
like NMR spin ensemble quantum computers or synthetic dimensions of 
ultra cold atoms for quantum simulations.
We here assume that we can obtain the signals as a macroscopic observable from the 
ensemble quantum system directly,
where the ensemble quantum system and the probe system are coupled
by an extremely weak interaction.
Actually, the NMR bulk ensemble average measurement is done in this way.
There is almost no backaction, or 
backaction is much smaller than other imperfections
like the T${}_1$ relaxation~\cite{NMRQC1,NMRQC2}.
In QRC, we make an active use of such a property of 
the ensemble quantum systems to exploit 
the complex quantum dynamics on the large degrees of freedom.

\subsection{Definition of quantum reservoir dynamics}
As nodes of the network of the QR, we use an orthogonal basis of quantum states.
The idea is similar to the quantum walks~\cite{Briegel14,QWalk2,QWalk1},
where each individual node is defined not by qubits (subsystems)
but by basis states like $\{|000\rangle, |001\rangle, ..., |111\rangle \}$.
Therefore, for $N$ qubits, we have $2^{N}$ basis states for a pure quantum state.
Moreover, here we employ the density matrix in general,
we define the nodes of the network by an orthogonal basis of the operator space
of the density matrices.
By using the Hilbert-Schmidt inner product, 
the density matrix can be represented as a vector $\mathbf{x}$ 
on a $4^N$-dimensional operator space.
Here the $i$-th coefficient $x_i$ of $\mathbf{x}$ 
is defined by $x_i = {\rm Tr}[B_i \rho]$ by using the set of $N$-qubit products of the Pauli operators $\{ B_i \}_{i=1}^{4^N} = \{ I, X, Y,Z\}^{\otimes N}$ (where $B_i B_j = \delta _{ij} I$).
Specifically, we choose the first $N$ elements such that $B_i =Z_i$ 
for convenience in the definition of the observables later.

In this operator space,
the time evolution is reformulated as a linear map
for the vector $\mathbf{x}$:
\begin{eqnarray}
\mathbf{x}(t+\tau) = U_{\tau} \mathbf{x}(t).
\end{eqnarray}
Here $U_{\tau}$ is a $4^N \times 4^N$ matrix whose element is defined by
\begin{eqnarray}
(U_{\tau})_{ji} := {\rm Tr}[ B_j e^{- i H \tau}B_i e^{ i H \tau}].
\end{eqnarray}
Owing to the unitarity of the dynamics $e^{- i H \tau} (e^{- i H \tau})^{\dag} = I$,
we have $U_{\tau} U_{\tau} ^{T}=I$. 
If the system is coupled to an external system for a measurement and/or a feedback operation, 
the time evolution (for the density matrix) is not given by the conjugation of the unitary operator 
$e^{- i H \tau}$; instead, it is generally given by a complete positive trace preserving (CPTP) map $\mathcal{D}$ for the density matrix $\rho$.
Even in such a case, the dynamics is linear, and hence the time evolution for $\mathbf{x}(t)$ is given in a linear form:
\begin{eqnarray}
\mathbf{x} \rightarrow W\mathbf{x}
\end{eqnarray}
where the matrix element is defined
\begin{eqnarray}
W_{ji} := {\rm Tr}[B_j \mathcal{D} (B_i)].
\label{eq:CPTPmap}
\end{eqnarray}

In order to exploit quantum dynamics for information processing, we have to introduce an input and the signals of the quantum system (see Fig.~\ref{figxx} (a)).
Suppose $\{ s_k \}_{k=1}^{M}$ is an input sequence, where $s_k$ can be a binary ($s_k \in \{0,1\}$) or a continuous variable ($s_k \in [0,1]$).
A temporal learning task here is to find, using the quantum system, 
a nonlinear function $y_k=f(\{ s_l \}_{l=1}^{k})$ such that 
the mean square error between $y_k$ and a target (teacher) output $\bar y_k$
for a given task becomes minimum.
To do so, at each time $t=k\tau$, the input signal $s_k$ is injected into a qubit, say the $1$st qubit, by replacing (or by using measurement and feedback) the $1$st qubit with the state $\rho _{s_k}= | \psi _{s_k}\rangle \langle \psi _{s_k} |$, where
\begin{eqnarray}
|\psi _{s_k} \rangle := \sqrt{1-s_k}|0\rangle + \sqrt{s_k} |1\rangle.
\end{eqnarray}
The density matrix $\rho$ of the system is transformed by the following 
CPTP map:
\begin{eqnarray}
\rho \rightarrow \rho _{s_k} \otimes {\rm Tr}_{1} [\rho ],
\end{eqnarray}
where ${\rm Tr}_{1}$ indicates the partial trace with respect to the first qubit.
The above action of the $k$th input on the state $\mathbf{x}(t)$ is again 
rewritten by a matrix $S_k$ by using Eq. (\ref{eq:CPTPmap}).
After the injection, the system evolves under the Hamiltonian $H$ 
for a time interval $\tau$.
Thus, the time evolution of the state for a unit timestep is given by
\begin{eqnarray}
\mathbf{x}(k\tau)=   U_{\tau} S_k \mathbf{x}\left((k-1)\tau \right).
\end{eqnarray}
After injecting the $k$th input,
the system evolves under the Hamiltonian 
for $\tau$ time. The time interval $\tau$
should be determined by both physical constraint for the input injections 
and performance of the QR.

The signal,
which is exploited for the learning process,
is defined as an average value of a local observable on each qubit.
We here employ, as observables, the Pauli operator $Z_i$ acting on each $i$th qubit.
For an appropriately ordered basis $\{B_i\}$ in the operator space,
the observed signals, and
the first $N$ elements of the state $\mathbf{x}(t)$
are related by $x_i(t) = {\rm Tr}[ Z_i \rho (t)]$ ($i=1,...,N$).
As we mentioned before, 
we do not consider the backaction of the measurements to obtain the average values $\{ x_i(t)\}$ 
by considering an ensemble quantum system.
We call the directly observed signals $\{ x_i(t)\}_{i=1}^{N}$ 
as the {\it true nodes}. 
Then, the remaining $(4^N-N)$ nodes of $\mathbf{x}(t)$ 
as {\it hidden nodes}, as they are not employed as the signals for learning.
For the learning,
we employ $x'_i(t)$ defined by
\begin{eqnarray}
x'_i(t) := {\rm Tr} [(I+Z_i)/2 \rho (t)] = (x_i (t)+1)/2 
\end{eqnarray}
by adding a constant bias and rescaling with $1/2$
just for a convenience for the presentation.

The unique feature of QRC in the reservoir computing context is that the exponentially many hidden nodes
originated from the exponentially large dimensions of the Hilbert space 
are monitored from a polynomial number of the signals defined as the true nodes 
as shown in Fig.~\ref{figxx} (b).
Contrast to a single quantum system, 
the ensemble quantum system allows us to make real-time use of the 
exponentially large degrees of freedom.
Note that at the injected one clean qubit at each time step and 
the single-qubit averaged outputs after a unitary time evolution
is enough hard for a classical computer to simulate efficiently in general~\cite{Fujiietal,FujiiMorimae}.

\subsection{Emerging nonlinearity from a linear system}
\begin{figure}
\centering
\includegraphics[width=90mm, clip]{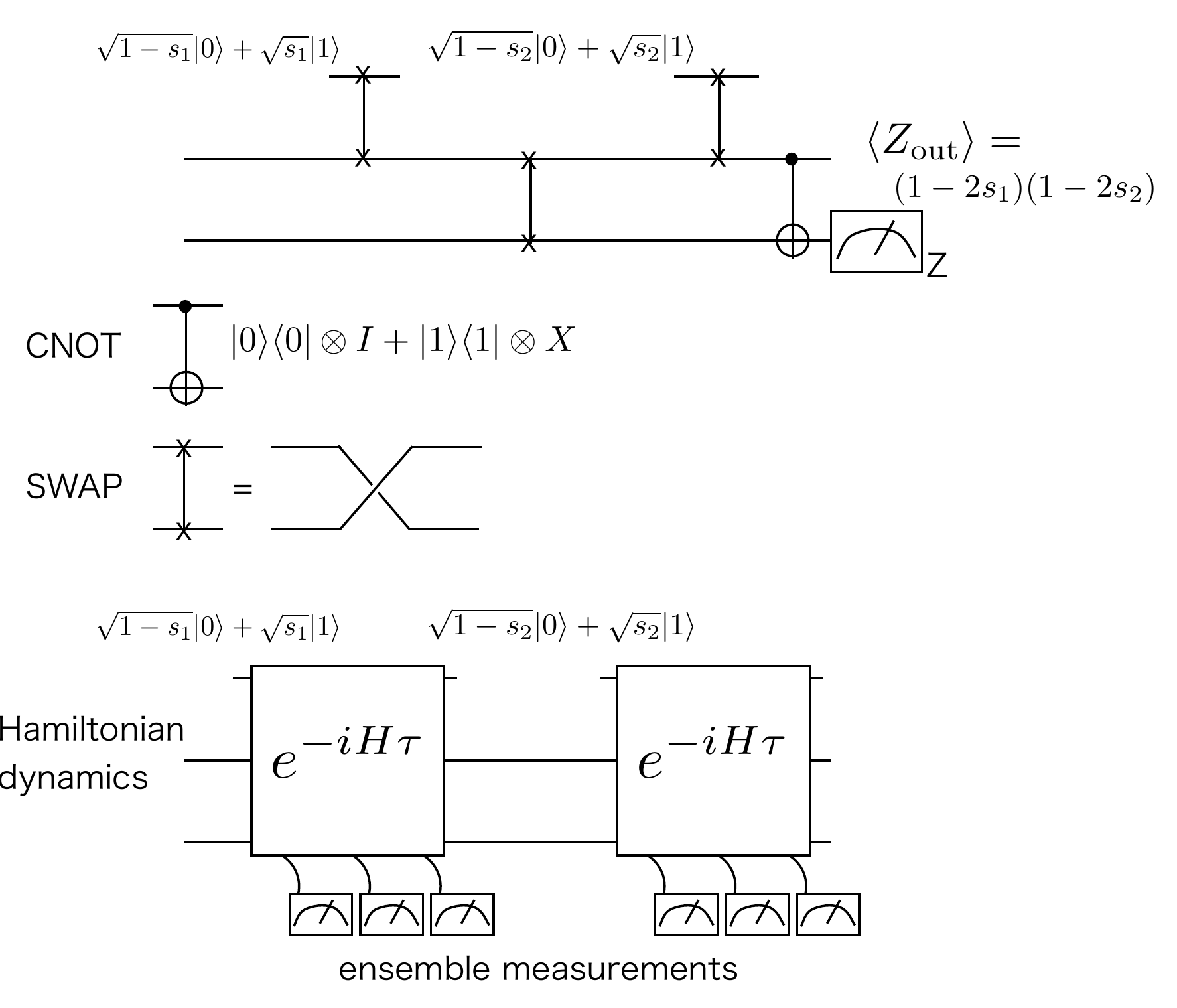}
\caption{Physical insight of QRC. (top) A quantum circuit 
whose output has the second order nonlinearity with respect to 
the input variables $s_1$ and $s_2$. (bottom) The quantum circuit is replaced by 
a unitary time evolution under a Hamiltonian $H$. The observables are 
monitored by the ensemble average measurements.
}
\label{figcircuit}
\end{figure}
We here provide a physical insight why quantum disordered dynamics
can be employed for nonlinear learning task.
One might think that the quantum system is totally linear, 
and hence that we cannot employ it for learning tasks, which essentially require nonlinearity.
However, this is not the case.
The definition of the nonlinearity defined for the learning task and 
the linearity of the dynamics on the quantum system are quite different.
Let us, for example, consider a quantum circuit shown in Fig.~\ref{figcircuit}.
For two input states $|\psi _{s_1}\rangle =\sqrt{1-s_1}|0\rangle + \sqrt{s_1} |1\rangle$
and $|\psi _{s_2}\rangle =\sqrt{1-s_2}|0\rangle + \sqrt{s_2} |1\rangle$,
we obtain $\langle Z_{\rm out}\rangle = (1-2s_1)(1-2s_2)$,
which has the second order nonlinearity with respect to $s_1$ and $s_2$.
Or equivalently, in the Heisenberg picture,
the observable $Z_{\rm out}$ corresponds to the nonlinear observable $Z_1 Z_2$.
Whereas dynamics is described as a linear map, information with respect to any kind of correlation exists in exponentially many degrees of freedom.
In the QRC,
such higher order correlations or nonlinear terms 
are mixed by the linear but quantum chaotic (non-integrable) dynamics $U_{\tau}$.
There exists a state corresponding to an observable $B_l = Z_i Z_j$, i.e. $x_{l}(t)={\rm Tr}[Z_iZ_j \rho(t)]$ storing correlation between $x_i(t)={\rm Tr}[Z_i \rho(t)]$ and $x_j (t)={\rm Tr}[Z_j \rho (t)]$, which can be monitored from another true node via $U_{\tau}$.
This mechanism allows 
us to find a nonlinear dynamics with respect to the input sequence $\{ s_k \}$ from the 
dynamics of the true nodes $\{ x_i(t) \} _{i=1}^{N}$.
The emergent nonlinearlity is not as special because classical (nonlinear) dynamics appears as (coarse-grained) dynamics of averaged values of the observables in the quantum system.
However, the universal emulation of nonlinear dynamics by training an optimal observable in disordered (chaotic) quantum systems explained below is unique for QRC, providing an alternative paradigm to digital universal quantum computing.

\subsection{Training readout weights}
\begin{figure}
\centering
\includegraphics[width=85mm, clip]{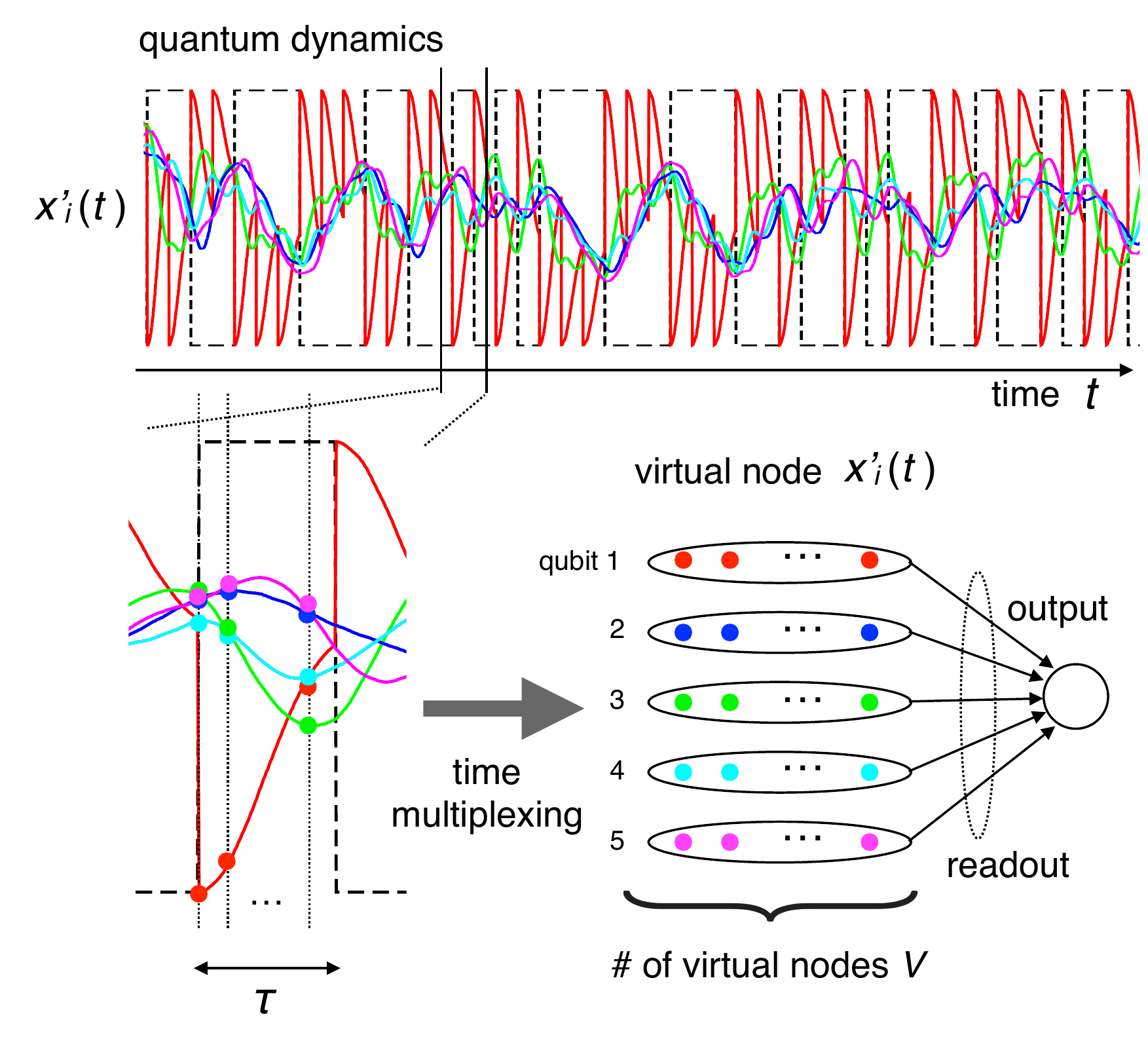}
\caption{Quantum reservoir dynamics and virtual nodes.
The time interval $\tau$ is divided into $V$ subdivided timesteps.
At each subdivided timestep the signals are sampled.
Using the $NV$ signals as the virtual nodes for each timestep $k$
in the learning phase,
the linear readout weights $\{ w_i^{\rm LR} \}$ are 
trained for a task.
 }
\label{fig1}
\end{figure}
Here we explain how to train the QR 
from the observed signals.
We harness complex quantum dynamics in a physically 
natural system by utilizing the reservoir computing approach.
Here the signals are sampled from the QR not only at the time $k \tau$, but also at each of the subdivided $V$ timesteps during the unitary evolution $U_{\tau}$ as shown in Fig.~\ref{fig1}.
That is,
at each time $t+v (\tau /V)$ with an integer $1\leq v \leq V$,
the signals $x_i'(t+v (\tau /V)) = {\rm Tr}[Z_i \rho (t+v (\tau /V))]$
are sampled.
Thus, at each timestep $k$, we have $NV$ virtual nodes in total.
These time multiplexed signals are denoted by $x'_{ki}$
with $i=n+vN$ with integers $1 \leq n \leq N$ and $0 \leq v \leq V$, $x'_{ki}$,
which means the signal of the $n$th qubit at time $t=k\tau + v(\tau/V)$, i.e. 
$x'_{ki}:=x'_i (k\tau + v(\tau/V))$.
We have named these {\it virtual nodes} (a similar technique 
of time multiplexing is also used in e.g. Ref.~\cite{Laser0}).
The virtual nodes allow us to make full use of the richness of quantum dynamics, because unitary real-time evolution is essential for nonlinearity.

Suppose learning is performed by using $L$ timesteps.
Let $\{x'_{ki}\}$ ($1\leq i \leq N V$ and $1\leq k \leq L$) be the states of the virtual nodes in the learning phase.
We also introduce $x'_{k0} =1.0$ as a constant bias term.
Let $\{\bar y_k\}_{k=1}^L$ be the target sequence for the learning.
In the reservoir computing approach, learning of a nonlinear function $y_m = f(\{ s_k \}_{k=1}^{m})$, which emulates the target sequence $\{\bar y_k \}$, is executed by training the linear readout weights 
of the reservoir states such that the mean square error 
\begin{eqnarray}
\frac{1}{L} \sum_{k=1}^{L}(y_k- \bar y_k)^2
\end{eqnarray} 
is minimised.
That is,
what we have to do is to find linear readout weights $\{ w_{i} \} _{i=0}^{N V}$ to obtain the output sequence
\begin{eqnarray}
y_k = \sum _{i=0}^{N V}x'_{ki}  w_{i} 
\end{eqnarray}
with the minimum mean square error.
This problem corresponds to solving the following equations:
\begin{eqnarray}
\bar{\mathbf{y}} = \mathbf{X} \mathbf{w},
\end{eqnarray}
where $\{x'_{ki}\}$, $\{\bar y_k\}_{k=1}^{L}$, and $\{w_i\}_{i=0}^{NV}$ are denoted by a $L \times (N V +1)$matrix $\mathbf{X}$, and column vectors $\bar{\mathbf{y}}$ and $\mathbf{w}$, respectively.
Here we assume that the length of the training sequence $L$ is much larger than the total number of the nodes $N V +1$ including the bias term.
Thus, the above equations are overdetermined, and hence the weights that minimise the mean square error are determined by the Moore-Penrose pseudo-inverse $\mathbf{X}^{+}:=(\mathbf{X}^{\rm T}\mathbf{X})^{-1}\mathbf{X}^{\rm T}$ ($(N V +1) \times L$ matrix) of $\mathbf{X}$ as follows:
\begin{eqnarray}
\mathbf{w}^{\rm LR} := \mathbf{X}^{+} \bar{\mathbf{y}}.
\end{eqnarray}
Using $\mathbf{w}^{\rm LR}$, we obtain the output from the QR 
\begin{eqnarray}
y_k = \sum _{i=0}^{NV} w^{\rm LR}_i  x'_{ki}.
\end{eqnarray}
Or equivalently, an optimal observable
\begin{eqnarray}
O_{\rm trained} \equiv \sum _{i=1}^{N}w^{\rm LR}_i (I+Z_i) /2 
+w^{\rm LR}_{N+1}  I
\end{eqnarray}
is trained, and the output is obtained as $\langle O_{\rm trained} \rangle$.

Specifically, as is the case in the conventional reservoir computing approach, none of the parameters of the system (Hamiltonian) requires fine tuning except for the linear readout weights.
Thus, we can employ any quantum system (Hamiltonian) as long as it exhibits dynamics with appropriate properties for our purpose, such as fading memory and nonlinearity.
That is, as long as the QR is sufficiently rich, we can find an optimal observable $O_{\rm trained}$ capable of exploiting the preferred behaviour via the training (learning) process.
In the following numerical experiments, we employ, as an example, the simplest quantum system, a fully connected transverse-field Ising model,
which exhibits a Wigner-Dyson statistics of the energy level spacing~\cite{Wigner,RMT,randomIsing}:
\begin{eqnarray} 
H = \sum _{ij} J_{ij} X_i X_j + h Z_i,
\label{eq_Ising}
\end{eqnarray}
where the coupling strengths are randomly chosen such that $J_{ij}$ is distributed randomly from $-J/2$ to $J/2$.
We introduce a scale factor $\Delta$ 
so as to make $\tau \Delta $ and $J/\Delta$ dimensionless.
Note that we do not employ any approximation,
but quantum dynamics of the above Hamiltonian is exactly calculated
to evaluate the potential performance of the QRs.
The imperfections including decoherence and noise on the observed signals, 
which might occur in actual experiments, 
are further taken into account in Sec.~\ref{sec:robustness}.

\section{Demonstrations of QRC for temporal learning tasks}
We start by providing several demonstrations to obtain a sense of QRC using a number of benchmark tasks in the context of machine learning. 
\begin{figure*}
\centering
\includegraphics[width=160mm, clip]{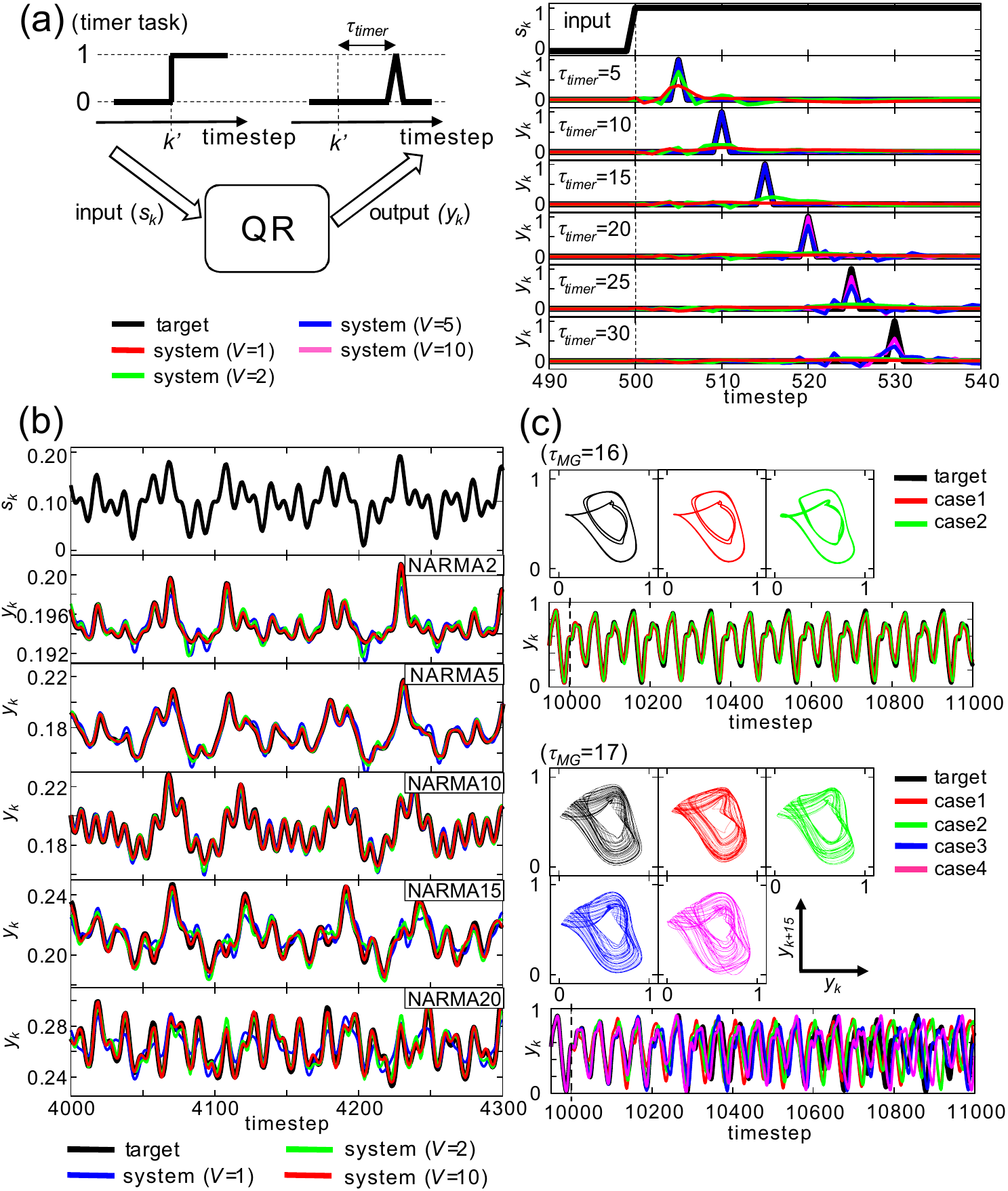}
\caption{Typical performances of QR for temporal machine learning tasks. 
(a) The timer task. 
A 6-qubit QR system is prepared, and starting from different initial conditions, 10 trials of numerical experiments were run for each $\tau_{\rm timer}$ setting.
$k'$ is set to $500$ throughout the numerical experiments.
The plots overlay the averaged system performance over 10 trials for $V=1, 2, 5,$ and $10$ with the target outputs.
(b) The NARMA emulation task.
This task requires five different NARMA systems driven by a common input stream to be emulated. The upper plot shows the input stream, and the corresponding task performances of a 6-qubit QR system for five NARMA tasks are plotted, overlaying the case for each $V$ with the target outputs.
(c) The Mackey-Glass prediction task.
The performances for $\tau_{\rm MG}=16$ (non-chaotic) and $17$ (chaotic) are shown.
The trained system outputs are switched to the autonomous phase at timestep 10000.
Two-dimensional plots, $(y_{k}, y_{k + 15})$, are depicted for the autonomous phase in each case.
For each setting of $\tau_{\rm MG}$, case 1 represents the case for 6 qubits, otherwise represent the cases for 7 qubits. 
For all tasks, the detailed settings and analyses are provided in Appendix.}
\label{fig2}
\end{figure*}

\subsection{Timer task}
Our first experiment is to construct a timer.
One important property of QRC is having memory to be exploited.
Whether the system contains memory or not can be straightforwardly evaluated by performing this timer task (see e.g., \cite{Buonomano1}).
The input is flipped from $0$ to $1$ at certain timestep ($k'$) as a cue, and the system should output $1$ if $\tau_{timer}$ timesteps have passed from the cue, otherwise it should output $0$ (see Fig.\ref{fig2} (a), left diagram).
To perform this task, the system has to be able to `recognize' the duration of time that has passed since the cue was launched.
This clearly requires memory. 
Here we used 6-qubit QRs with $\tau \Delta=1$ to perform this task by incrementally varying $V$.

Figure \ref{fig2} (a) shows the task performance with trained readouts.
We can clearly observe that by increasing $V$ the performance improved, which means that the amount of memory, which can be exploited, also increased.
In particular, when $V=5$ and $10$, the system outputs overlap with the target outputs within the certain delay, which clearly demonstrates that our QR system is capable of embedding a timer.
By increasing the delay timesteps $\tau_{\rm timer}$, we can gradually see that the performance declines, which expresses the limitation of the amount of memory that can be exploited within the QR dynamics.
It is interesting to note that while the systems are highly disordered, we can find an observable $O_{\rm trained}$ or a mode, at which the wave function of the system is focused after a desired delay time $\tau _{\rm timer}$.
This is very useful as a control scheme for engineering quantum many-body dynamics.
For further information, see detailed settings, experimental and learning procedures, and analyses for the timer task in Appendix~\ref{Ap:timer}.

\subsection{NARMA task}
The second task is the emulation of nonlinear dynamical systems, called {\it nonlinear auto-regressive moving average} (NARMA) systems, which is a standard benchmark task in the context of recurrent neural network learning. 
This task presents a challenging problem for any computational system because of its nonlinearity and dependence on long time lags \cite{long_short}.
The first NARMA system is the following second-order nonlinear dynamical system:
\begin{equation}
y_{k+1} = 0.4y_{k} + 0.4y_{k}y_{k-1} + 0.6s^{3}_{k} + 0.1.
\end{equation}
This system was introduced in Ref.~\cite{benchmark} and used, for example, in Refs.~\cite{Kohei1,Kohei3}. 
For descriptive purposes, we call this system NARMA2.
The second NARMA system is the following nonlinear dynamical system that has an order of $n$:
\begin{equation}
y_{k+1} = \alpha y_{k} + \beta y_{k} (\sum_{j=0}^{n-1}y_{k-j}) + \gamma s_{k-n+1}s_{k} + \delta,
\end{equation}
where $(\alpha, \beta, \gamma, \delta)$ are set to $(0.3, 0.05, 1.5, 0.1)$, respectively.
Here, $n$ is varied using the values of $5, 10, 15,$ and $20$, and the corresponding systems are called NARMA5, NARMA10, NARMA15, and NARMA20, respectively.
In particular, NARMA10 with this parameter setting was introduced in Ref.~\cite{benchmark} and broadly used (see, e.g., Refs.~\cite{Reservoir,Kohei1,Kohei3}).
As a demonstration, the input $s_{k}$ is expressed as a product of three sinusoidal functions with different frequencies. 
(Note that when the input is projected to the first qubit, the value is linearly scaled to $[0, 1]$; see Appendix~\ref{ap:NARMA} for details).
Here, according to an input stream expressed as a product of three sinusoidal functions with different frequencies, the system should simultaneously emulate five NARMA systems (NARMA2, NARMA5, NARMA10, NARMA15, and NARMA20), which we call multitasking.

Figure \ref{fig2} (b) plots the input sequence and the corresponding task performance of our 6-qubit QR system with $\tau \Delta=1$ with trained readout by varying $V$.
We can clearly observe that by increasing $V$, the performance improves, so that when $V=10$, the system outputs almost overlap with the target outputs.
Further information and extended analyses on the tasks with random input streams can be found in Appendix~\ref{ap:NARMA}.

\subsection{Mackey-Glass prediction task}
The third experiment is a Mackey-Glass (MG) time series prediction task, including a chaotic time series.
This is also a popular benchmark task in machine learning (e.g., \cite{Jaeger0}).
Here, unlike the previous two cases, the system output is fed back as the input for the next timestep, which means that when the system with trained readout generates outputs, it receives its own output signals through the feedback connections instead of through external inputs.
To train the readout weights, the system is forced by the correct teacher output during presentation of the training data, without closing the loop.
A slight amount of white noise is added to the reservoir states in the training phase to make the trained system robust, and the weights are trained through the usual procedure (see Appendix~\ref{Ap:MG} for further information).
The MG system has a delay term $\tau_{\rm MG}$, and when $\tau_{\rm MG} > 16.8$ it exhibits a chaotic attractor.
We first test a non-chaotic case ($\tau_{\rm MG} = 16$) for comparisons and then test the chaotic case, where $\tau_{\rm MG} = 17$, which is the standard value employed in most of the MG system prediction literature.

Figure \ref{fig2} (c) depicts the typical task performances of 6- and 7-qubit QR systems.
When $\tau_{\rm MG} = 16$, the system outputs overlap the target outputs, which implies successful emulations.
When $\tau_{\rm MG} = 17$, our systems tend to remain relatively stable in the desired trajectory for about 200 steps, after switched from teacher forced condition, start to deviate perceptibly large.
Furthermore, checking a two-dimensional plot by plotting points $(y_{k}, y_{k + 15})$, it appears that the learned model has captured the essential structure of the original attractor (e.g., when $\tau_{\rm MG} = 17$, the model actually demonstrates chaos). 
In both tasks, the 7-qubit QR systems generally performed better than the 6-qubit QR systems.
Further details can be found in Appendix~\ref{Ap:MG}.

\section{Performance analyses}
\begin{figure*}
\centering
\includegraphics[width=150mm, clip]{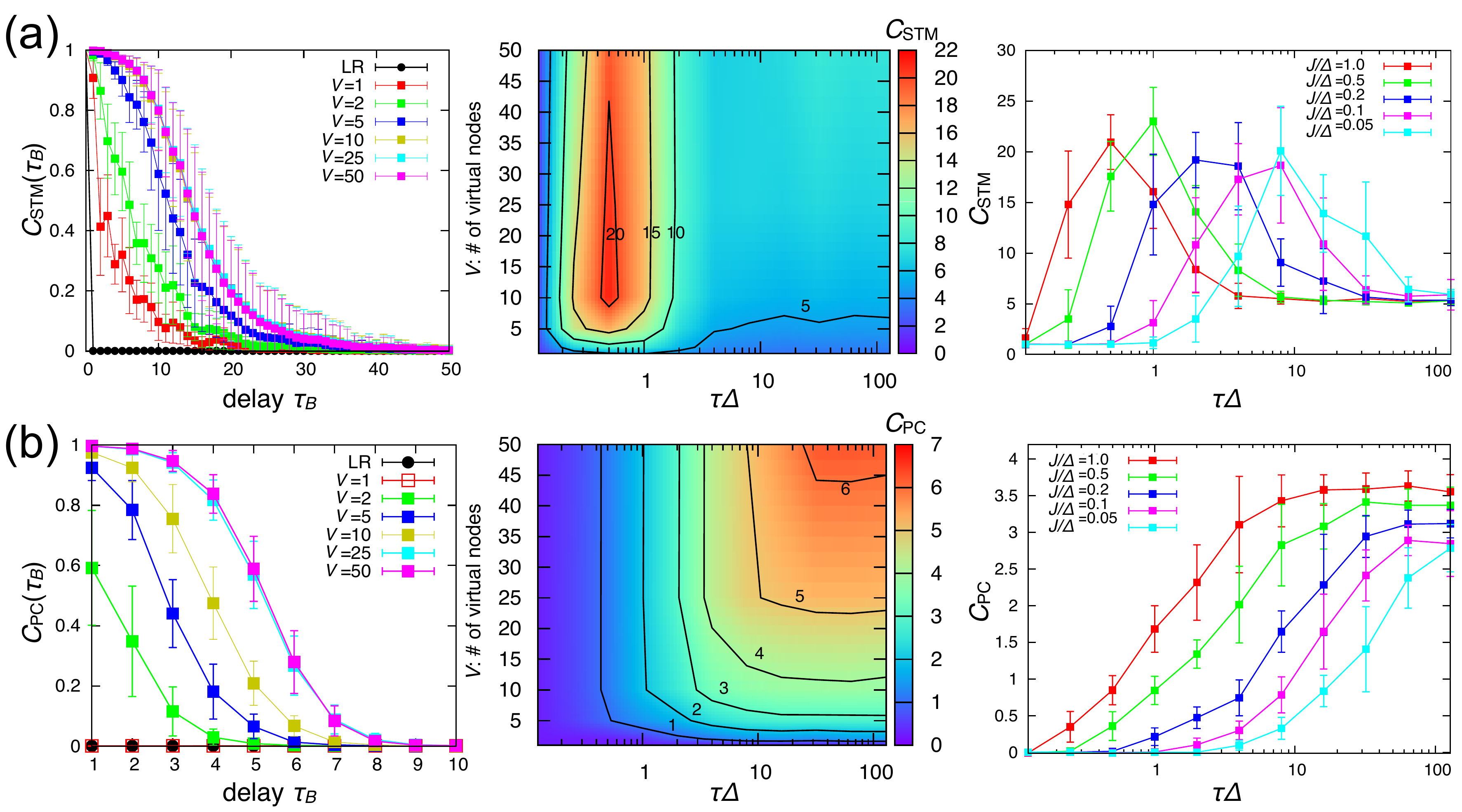}
\caption{Performance analyses of the 5-qubit QRs.
(a) (left) STM curve $C_{\rm STM} (\tau_{B})$ plotted as a function of the delay $\tau_{B}$ 
for $J/\Delta=2h/\Delta=1$, $\tau\Delta=1$ and $V=1$--$50$.  
(middle) STM capacity $C_{\rm STM}$ plotted 
as a function of the number of virtual nodes $V$ 
with the same QR settings for $\tau \Delta= 0.5$--$128$. 
(right) STM capacity $C_{\rm STM}$ plotted 
as a function of $\tau \Delta$ with couplings
$J/\Delta=0.05$--$1.0$ and $h/\Delta = 0.5$.
(b) PC curve $C_{\rm PC} (\tau_{B})$ and capacity $C_{\rm PC}$ 
plotted with the same settings as (a).
The error bars in the left and right panels indicate
the standard deviations of the capacities evaluated on 20 samples 
of the QRs with respect to the random couplings.
}
\label{fig3}
\end{figure*}
We perform detailed analyses on the computational capabilities of the 5-qubit QRs focusing on the two popular benchmark tasks of Boolean function emulations over a binary input sequence (see e.g., Refs. \cite{Kohei2,EdgeofChaos}), which we name the {\it short-term memory} (STM) task and {\it parity check} (PC) task. 
The former task is intended to emulate a function that outputs a version of the input stream delayed by $\tau_{B}$ timesteps, whereas the latter is intended to emulate an $\tau_{B}$-bit parity checker. 
Both tasks require memory to be emulated, and the PC task requires nonlinearity in addition, because the parity checker function performs nonlinear mapping. 
Hence, the STM task can evaluate the memory capacity of systems and the PC task can additionally evaluate the amount of nonlinearity within systems. 

The function for the STM task can be expressed as follows:
\begin{equation*}
y_{k} = s_{k-\tau_{B}},
\end{equation*}
where $s_{k}$ is a binary sequence and $\tau_{B}$ represents the delay.
The function for the PC task is expressed as follows:
\begin{align*}
y_{k} &= Q(\sum_{m=0}^{\tau_{B}}s_{k-m}), \\
Q (x) &= \left \{
\begin{array}{l}
0 \ \ (x \equiv 0 \ ({\rm mod} \ 2)) \\
1 \ \ ({\rm otherwise}).
\end{array}
\right.
\end{align*}
We investigated both tasks thoroughly by applying a random input sequence for the tasks such that there is no external source to provide temporal coherence to the system.
In these tasks, one trial consists of 5000 timesteps.
The first 1000 timesteps are discarded, the next 3000 timesteps are used for training, and the last 1000 timesteps are used for system evaluation. 
We evaluated the system performance with the target output for each given $\tau_{B}$ by using the measure known as $\tau_{B}$-delay capacity $C(\tau_{B})$ expressed as
\begin{equation*}
C(\tau_{B}) = \frac{{\rm cov}^{2}(y_{k}, \bar{y}_{k})}{\sigma^{2}(y_{k})\sigma^{2}(\bar{y}_{k})}.
\end{equation*}
In the main text, $\tau_{B}$-delay capacities for the STM task and the PC task are termed $\tau_{B}$-delay STM capacity $C_{STM}(\tau_{B})$ and $\tau_{B}$-delay PC capacity $C_{PC}(\tau_{B})$, respectively.
Note that, in the analyses, to reduce a bias due to the effect of the finite data length, we have subtracted $C(\tau_{B}^{\rm max})$ from $C(\tau_{B})$, where $\tau_{B}^{\rm max}$ is a substantially long delay.
The capacity $C$ is defined as 
\begin{equation*}
C = \sum_{\tau_{B} = 0}^{\tau_{B}^{\rm max}} C(\tau_{B}),
\end{equation*}
where $\tau_{B}^{\rm max}$ was 500 throughout our experiments. 
The capacities for the STM task and the PC task are referred to as the STM capacity $C_{STM}$ and the PC capacity $C_{PC}$, respectively.
For each task, 20 samples of the QRs were randomly generated, and the average values of the $\tau_{B}$-delay capacities and the capacities were obtained.    

In Fig.~\ref{fig3} (a) (left), $C_{\rm STM} (\tau_{B})$ is plotted as a function of $\tau_{B}$ for $V=1,...,50$, where $\tau \Delta=1 $ and $J/\Delta=1.0$ are set.
The abrupt decay exhibited by the curve 
is improved when the number of virtual nodes is increased.
In Fig.~\ref{fig3} (a) (middle), the STM capacity is plotted as a function of the number of virtual nodes $V$ and the time interval $\tau \Delta$.
It shows that the STM capacity becomes saturated around $V=10$.
The 5-qubit QRs with $\tau \Delta = 0.5$ and $1.0$ exhibit a substantially high STM capacity $\sim 20$, which is much higher than that of the ESNs of 500 nodes (see Sec.~\ref{sec:character} for details).
A plot of the STM capacity as a function of $\tau$ for a fixed number of virtual nodes $V=10$ does not exhibit monotonic behaviour as shown in Fig.~\ref{fig3} (right).
This behaviour is understood as follows.
In the limit of $\tau \rightarrow 0$,
the dynamics approach an identity map and hence
become less attractive,
and this is more desirable to maintain the separation among different inputs.
At the same time, a shorter $\tau$ implies 
less information is embedded in the present input setting.
In the limit of larger $\tau$, on the other hand,
the input sequence is injected effectively;
however, the dynamics become attractive, and 
the separation fades rapidly.
Originating from these two competing effects,
there is an optimal time interval $\tau$ for which the STM capacity is maximised.

In Fig.~\ref{fig3} (b) (left), $C_{\rm PC} (\tau_{B})$ is plotted as a function of $\tau_{B}$ for $V=1,...,50$.
Specifically, $C_{\rm PC} (\tau_{B})$ is exactly zero when $V=1$.
This clearly shows that 
the virtual nodes, which spatialize the real-time dynamics during the interval $\tau$, are important to extract nonlinearity.
In Fig.~\ref{fig3} (b) (middle), 
the PC capacity is plotted as a function of the number of virtual nodes $V$ and 
the time interval $\tau \Delta$.
As expected, the longer the time interval $\tau$ is, the higher the PC capacity exhibited by the QR,
as shown in Fig.~\ref{fig3} (middle and right).
This is because the true nodes are able to increase communication with the virtual nodes.
The number of virtual nodes required 
for the saturation of the PC capacity is also increased in the case of a longer $\tau$.

\subsection{Characterizations of QRs}
\label{sec:character}
\begin{figure}
\centering
\includegraphics[width=80mm, clip]{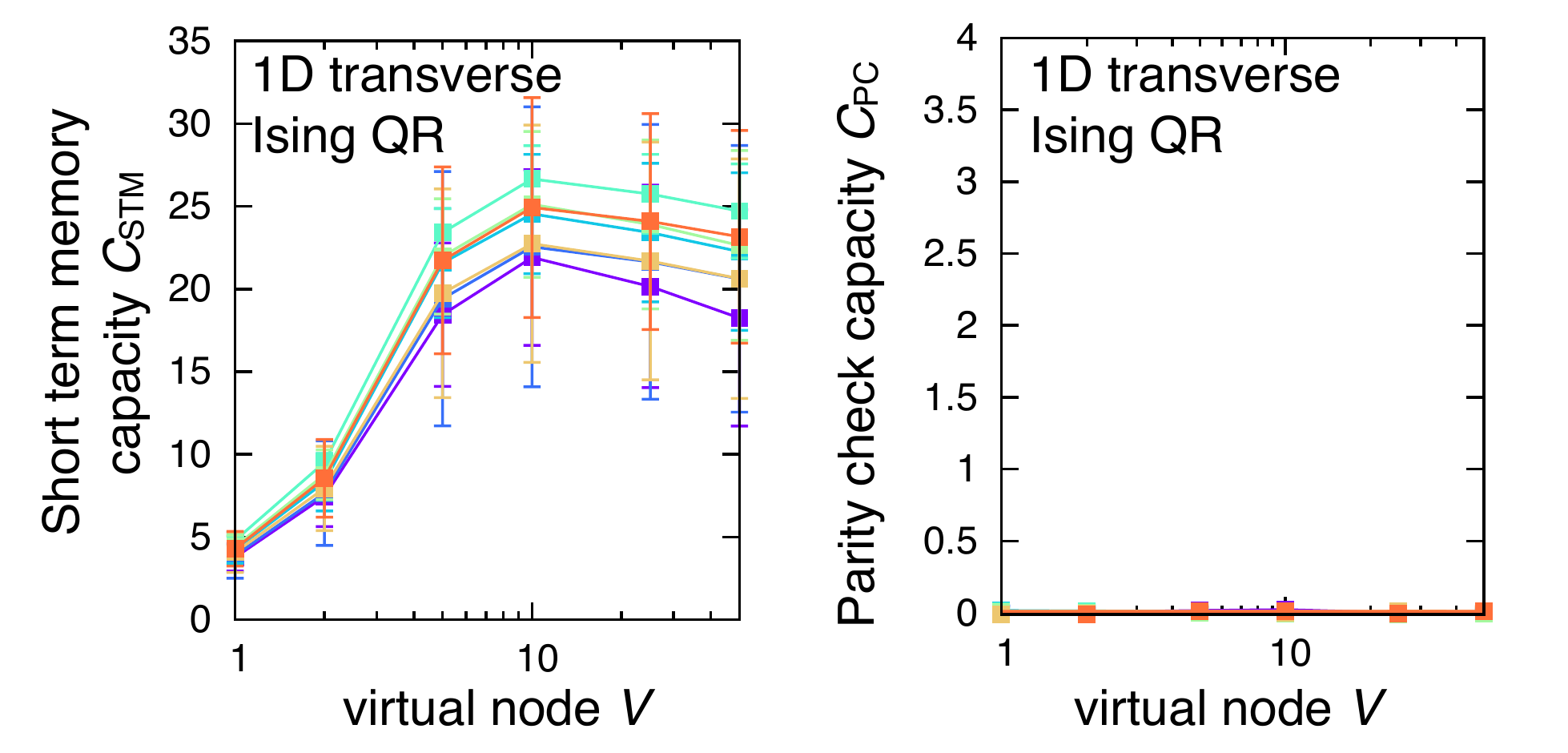}
\caption{
STM (left) and PC (right) capacities for the 1D transversal Ising model.
The error bars show the standard deviations evaluated 
on 20 samples of the QRs with respect to the random couplings}
\label{figSI7}
\end{figure}
\begin{figure*}[t]
\centering
\includegraphics[width=160mm, clip]{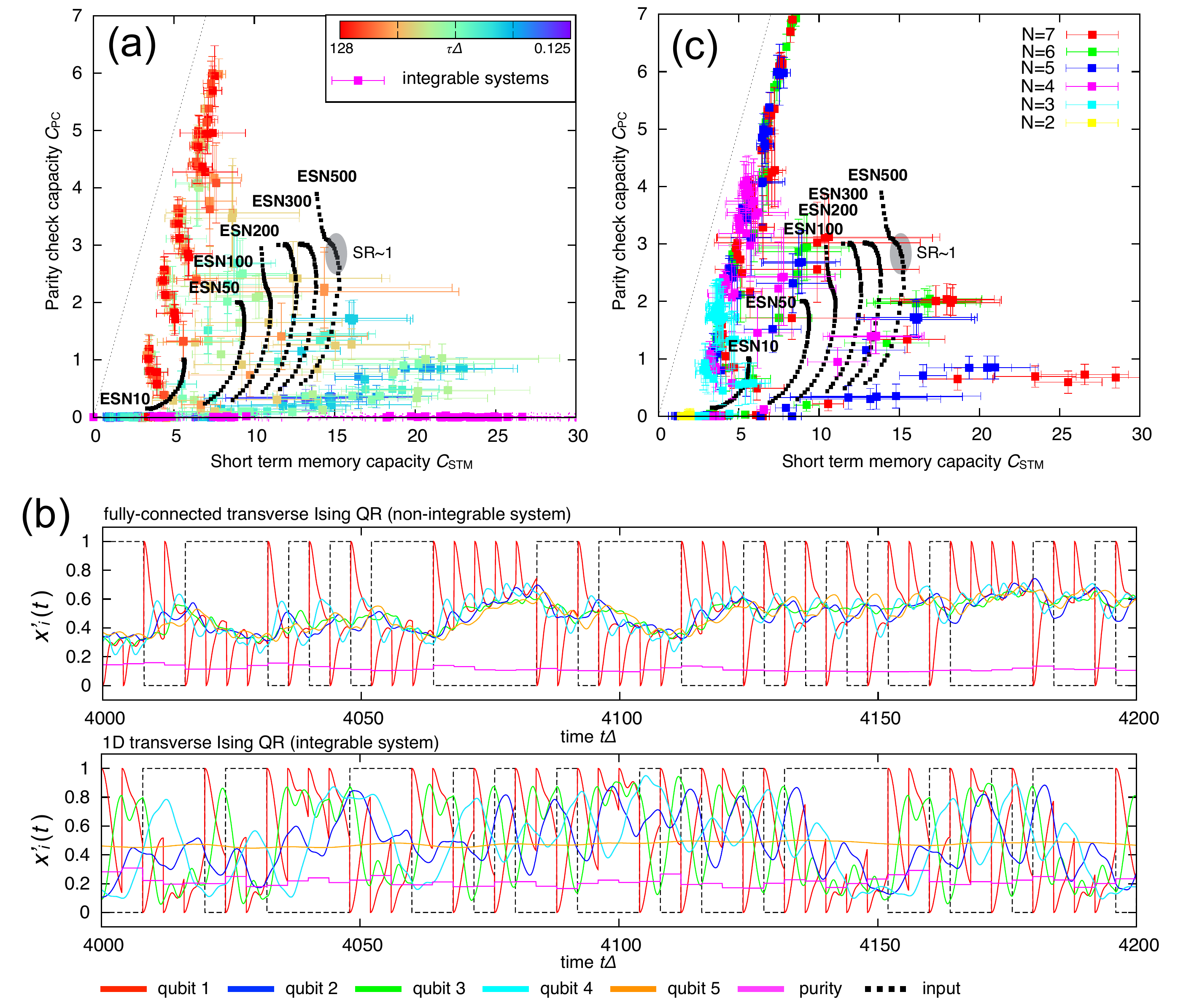}
\caption{STM and PC capacities  
under various settings. (a) Capacities for the 5-qubit QRs 
plotted with various parameters 
$\tau \Delta = 0.125$--$128$, 
$J/\Delta=0.05$--$1.0$, $h/\Delta = 0.5$,
and $V=1$--$50$.
Integrable cases with 1DNN couplings
are shown as ``integrable systems".
For each setting, the capacities are 
evaluated as an average on 20 samples, and the standard deviations are shown 
by the error bars.
(b) Typical dynamics with fully connected (upper)
and 1DNN couplings (lower) are shown with the signals 
from each qubit.
The input sequence and purity,
a measure of quantum coherence, are shown by dotted and solid lines, respectively.
(c) QRs with $J/\Delta=2h/\Delta =1$ and $
0.125 \leq \tau\Delta  \leq 128$ for $N=2$--$7$.
For each setting, the capacities are 
evaluated as an average on 20 samples, and the standard deviations are shown 
by the error bars.
The ESNs with 10--500 nodes 
are shown as references.}
\label{fig4}
\end{figure*}

Let us clarify the unique properties of the QRs in terms of the STM and PC capacities.
We plot $(C_{\rm STM}, C_{\rm PC})$ 
for the 5-qubit QRs with various coupling settings in Fig.~\ref{fig4} (a),
which include 
a restricted type of QR with one-dimensional nearest-neighbour (1DNN) couplings,
i.e. $J_{ij}\neq 0$ only for $j=i+1$ in Eq. (\ref{eq_Ising}).
In this case, the transversal-field Ising model becomes integrable, that is, 
exactly solvable by mapping it into a free-fermionic model
via the Jordan-Wigner transformation.
Because the effective dimension of the state space
is reduced from $2^{2N}$ to $2N$,
the amplitudes of the oscillations 
are larger for the 1DNN case as shown in Fig. ~\ref{fig4} (b).
From the real-time dynamics,
one might expect a rich computational capability
even for the integrable dynamics.
Although this is true for the STM capacity, 
it does not hold for the PC capacity.
As shown in Fig.~\ref{figSI7},
the STM capacity of the 1DNN QRs is extremely high
above 20. However, the PC capacity 
is substantially poor, which cannot 
improve even if the time intervals $\tau$ or 
the number of virtual nodes are changed.
This is a natural consequence of 
the inability of the 1DNN model to fully employ 
exponentially large state spaces.
In this way,
the computational capacity of QRs,
especially their nonlinear capacity, 
has a close connection with the 
nonintegrability of the underling QR dynamics.
This implies that the computational capacity 
as a QR provides a good metric of the integrability of 
quantum dynamics.
A nonintegrable quantum system 
is identified as quantum chaos,
which is specified by the Wigner-Dyson distribution of the 
energy eigenstate spacing.
The operational metric of the integrability of quantum dynamics
would be useful to build a modern operational understanding of 
quantum chaos by relating it 
to the emulatability of classical chaos.

\begin{figure*}[t]
\centering
\includegraphics[width=150mm, clip]{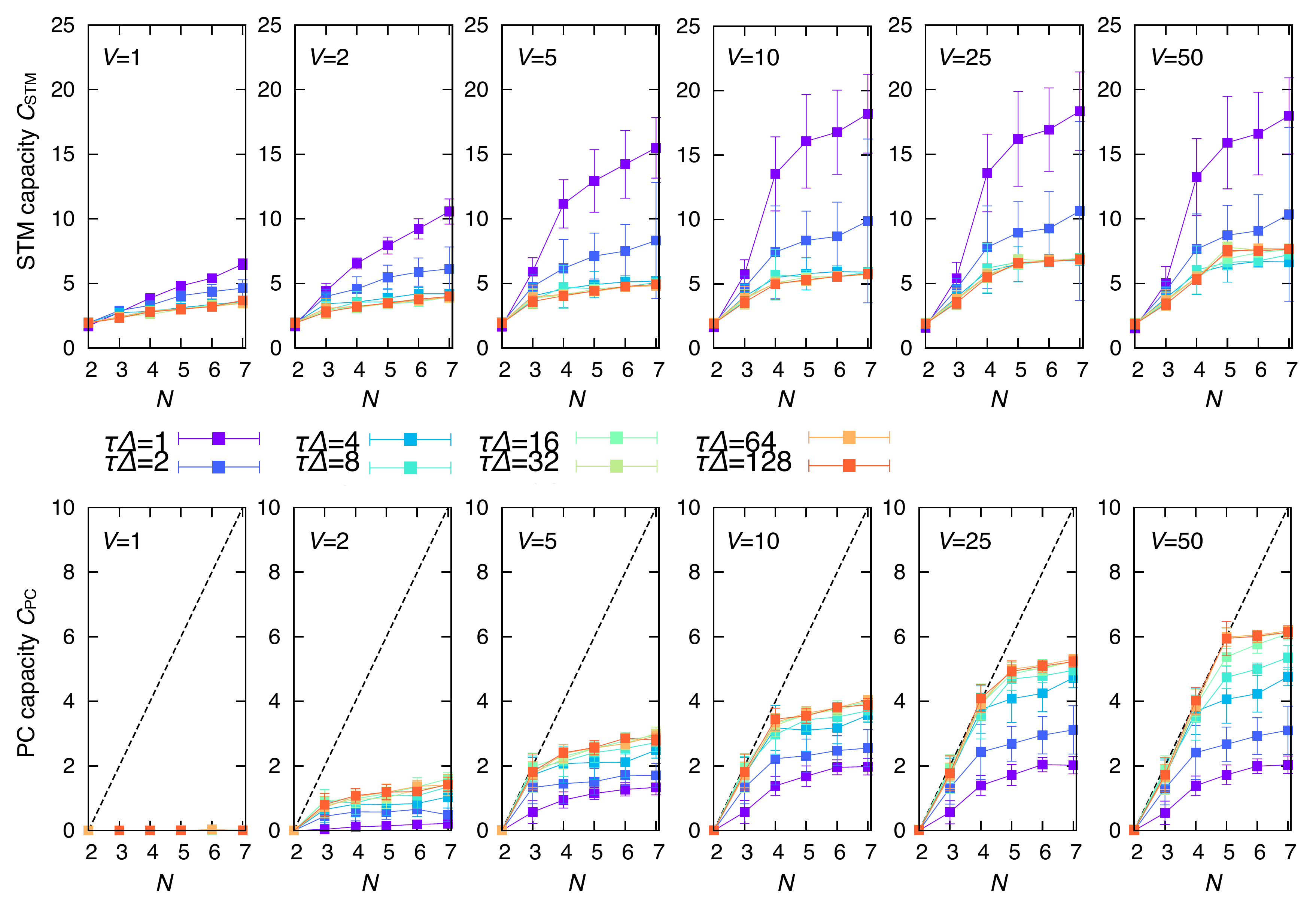}
\caption{Scaling of the STM and PC capacities against the number of the qubits.
(Top) The STM capacity $C_{\rm STM}$ 
is plotted against the number of qubits $N$
for each number of virtual nodes ($V=1,2,5,10,25,50$ from left to right).
(Bottom) The PC capacity $C_{\rm PC}$ is plotted against the number of 
qubits $N$ for each number of virtual nodes ($V=1,2,5,10,25,50$ from left to right). $C_{\rm PC}=2(N-2)$ is shown by dotted lines.
The error bars show the standard deviations evaluated 
on 20 samples of the QRs with respect to the random couplings.}
\label{figSI4}
\end{figure*}

Next we investigate the 
scaling of the STM and PC capacities against 
the number of the qubits $N$ in the QRs. 
In Fig.~\ref{figSI4}, 
the STM and PC capacities are plotted 
against the number of qubits
for the virtual nodes $V=1,2,5,10,25,$ and $50$.
First, 
both capacities monotonically increase 
in the number of the qubits $N$ and the virtual nodes $V$.
Thus, by increasing the time resolution and size of the QR,
we can enhance its computational capability.
The STM capacity is improved by 
increasing the number of virtual nodes $V$
especially for optimally chosen time intervals $\tau$.
The improvement saturates around $V=10$.
The scaling behaviour of the STM capacity 
seems to be different for $N=2$--$4$ and $N=4$--$7$
when the virtual nodes are introduced.
For optimally chosen time intervals, 
the STM capacity seems to increase linearly 
in terms of the number of qubits.

The PC capacity also increases 
in terms of the number of virtual nodes $V$, but 
its saturation highly depends on the choice of the time interval $\tau$.
For a short interval $\tau\Delta=1$, the PC capacity 
saturates around $V=10$. However, for $\tau \Delta =128$,
it seems not to saturate even with $V=50$.
In any case, the PC capacity seems to 
increase linearly in terms of the number of the qubits $N$.
Interestingly, at the 
large $\tau$ and large $V$ limits,
the PC capacity saturates the line defined by $C_{\rm PC} = 2(N-2)$.
The origin of this behaviour is completely unknown at this moment.

In Fig.~\ref{fig4} (c), the STM and PC capacities are plotted for the QRs from $N=2$ to $N=7$.
The 7-qubit QRs, for example, with $\tau \Delta=2$, $J/\Delta =2h/\Delta=1$, and $V=10$--$50$, 
are as powerful as the ESNs of 500 nodes with the spectral radius 
around 1.0. Note that even if the virtual nodes are included,
the total number of nodes $NV=350$ is less than 500.

\subsection{Robustness against imperfections}
\label{sec:robustness}
We here investigate the effect of decoherence (noise) to validate the
feasibility of QRC. 
We consider two types of noise:
the first is decoherence, which is introduced by 
an undesired coupling of QRs with the environment,
thereby resulting in a loss of quantum coherence,
and the other is a statistical error 
on the observed signals from QRs.
The former is more serious because 
quantum coherence is, in general, 
fragile against decoherence, 
which is the most difficult barrier for 
realizations of quantum information processing.

\begin{figure*}[t]
\centering
\includegraphics[width=170mm, clip]{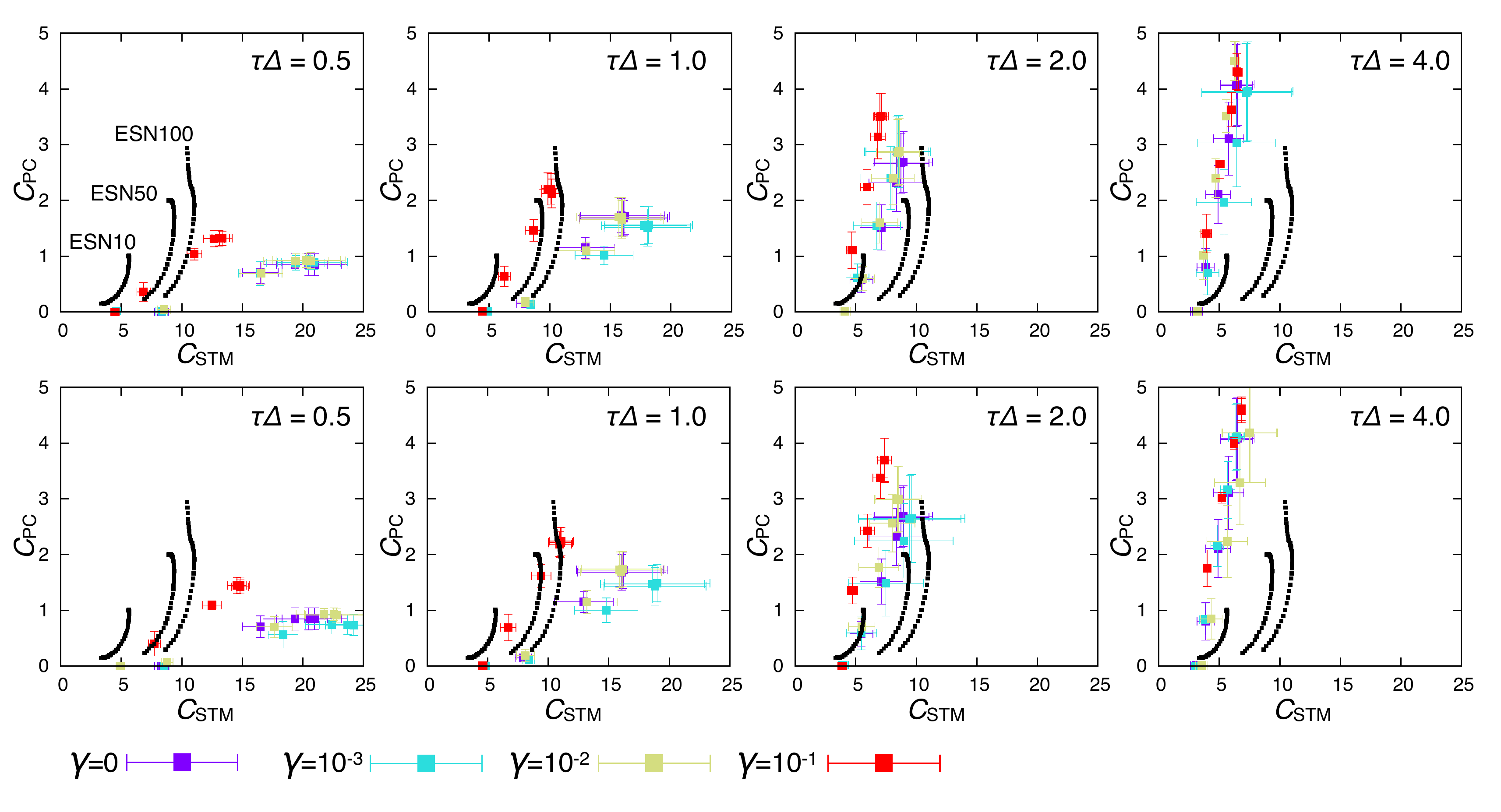}
\caption{STM and PC capacities under decoherence investigated for the 5-qubit QRs.
The parameters are set as $\tau \Delta = 0.5, 1.0, 2.0, 4.0$ and $V=1,2,5,10,25,50$.
(Top) Capacities $(C_{\rm STM}, C_{\rm PC})$ under dephasing in the $z$-axis are plotted for $\gamma =10^{-2},10^{-3}, 10^{-4}$. 
(Bottom) Capacities $(C_{\rm STM}, C_{\rm PC})$ under dephasing in the $x$-axis are plotted for $\gamma =10^{-2},10^{-3}, 10^{-4}$. The error bars show the standard deviations evaluated 
on 20 samples of the QRs with respect to the random couplings.}
\label{figSI5}
\end{figure*}
We employ the dephasing noise as decoherence,
which is a simple yet experimentally dominant 
source of noise.
In the numerical simulation,
the time evolution is divided into 
a small discrete interval $\delta t$,
and qubits are exposed to the 
single-qubit phase-flip channel
with probability $(1-e^{-2 \gamma \delta t})/2$ for each timestep:
\begin{eqnarray}
\mathcal{E}(\rho) = 
\frac{1+e^{-2 \gamma \delta t}}{2}\rho + \frac{1-e^{-2 \gamma \delta t}}{2}Z\rho Z.
\end{eqnarray}
This corresponds to a Markovian dephasing 
with a dephasing rate $\gamma$ and 
destroys quantum coherence, i.e. 
off-diagonal elements in the density matrix.
Apart from the dephasing in the $z$-direction,
we also investigate the dephasing in $x$-direction,
where the Pauli $Z$ operator is replaced by $X$.
In Fig.~\ref{figSI5},
the STM and PC capacities $(C_{\rm STM}, C_{\rm PC})$
are plotted for $\tau \Delta = 0.5,1.0,2.0,$ and $4.0$ (from left to right)
with $V=1,2,5,10,25,$ and $50$ and $\gamma = 10^{-1}, 10^{-2},$ and $10^{-3}$.
The results show that dephasing of the rates $10^{-2}-10^{-3}$,
which is within an experimentally feasible range,
does not degrade computational capabilities.
A subsequent increase in the dephasing rate
causes the STM capacity to become smaller,
especially for the case with a shorter time interval $\tau \Delta =0.5$.
On the other hand, 
the PC capacity is improved by increasing 
the dephasing rate.
This behaviour can be understood as follows.
The origin of quantum decoherence is 
the coupling with the untouchable 
environmental degree of freedom,
which is referred to as a ``reservoir" 
in the context of open quantum systems.
Thus, decoherence 
implies an introduction of another dynamics 
with the degree of freedom in the ``reservoir" computing framework.
This leads to the decoherence-enhanced improvement of 
nonlinearity observed in Fig.~\ref{figSI5},
especially for a shorter $\tau$ with less rich dynamics.
Of course, for a large decoherence limit, 
the system becomes classical, preventing us from 
fully exploiting the potential computational capability of the QRs.
This appears in the degradation of the STM capacity.
By attaching the environmental degree of freedom,
the spatialized temporal information is more likely 
to leak outside the true nodes.
Accordingly we cannot reconstruct a past input sequence 
from the signals of the true nodes.
In other words, 
quantum coherence is important to 
retain information of the past input sequence
within the addressable degree of freedom.
In short, 
in the QRC approach, 
we do not need to distinguish between coherent dynamics and decoherence;
we can exploit any dynamics on the quantum system as it is,
which is monitored only from the addressable degree of freedom 
of the quantum system.

\begin{figure}[t]
\centering
\includegraphics[width=80mm, clip]{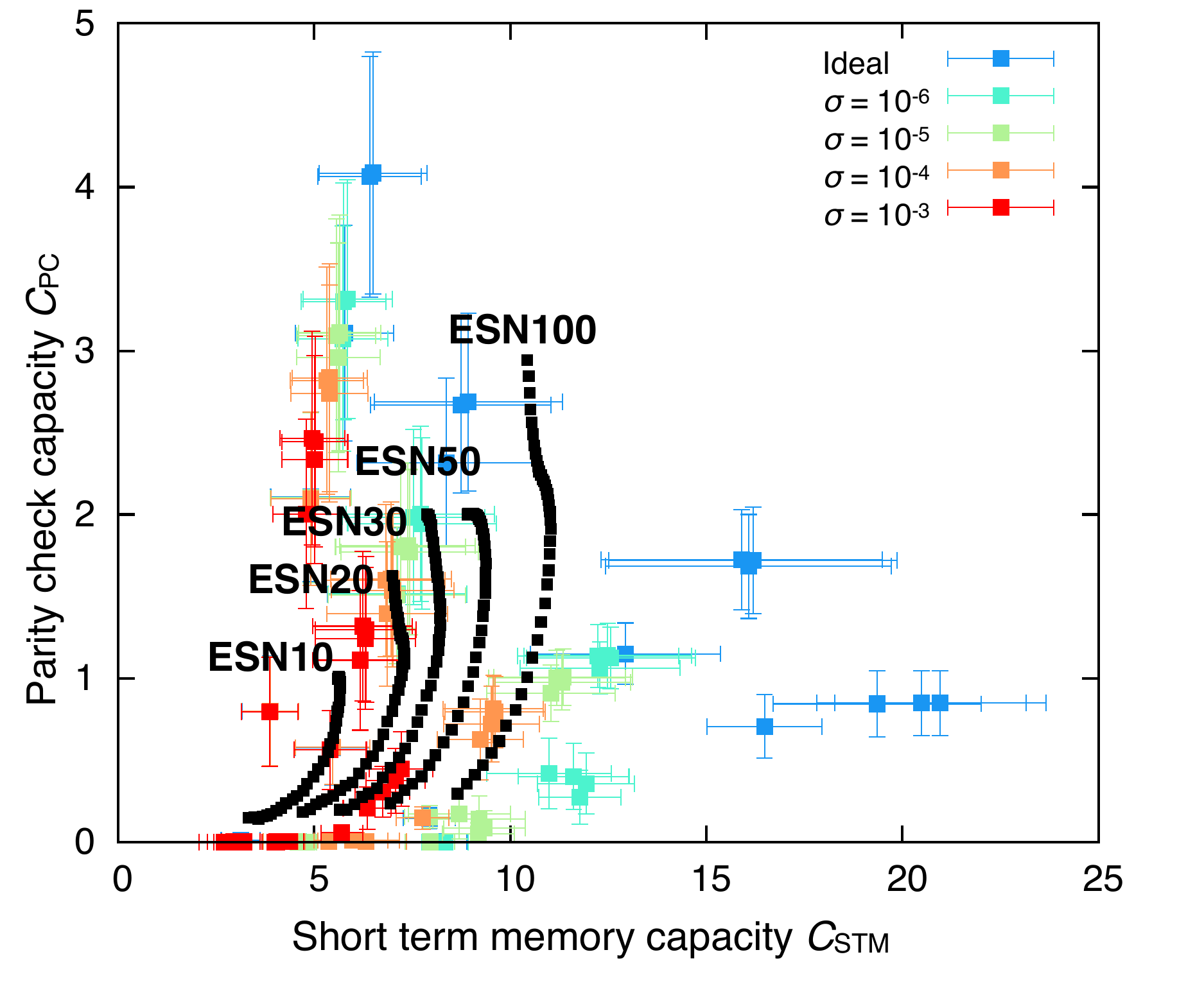}
\caption{Effect of the statistical error on the observed signals investigated for the 5-qubit QRs.
The parameters are set as $\tau \Delta = 0.5, 1.0, 2.0, 4.0$ and $V=1,2,5,10,25,50$. 
The STM and PC capacities $(C_{\rm STM},C_{\rm PC})$ are plotted 
for Gaussian noise with zero mean and variance $\sigma = 10^{-3},10^{-4},10^{-5},10^{-6}$. The error bars show the standard deviations evaluated 
on 20 samples of the QRs with respect to the random couplings. The performances 
for the ESNs are calculated without adding the observational noise.}
\label{figSI6}
\end{figure}

Next, we consider the statistical noise 
on the observed signal from the QRs.
We investigate the STM and PC capacities 
by introducing Gaussian noise with zero mean and variance $\sigma$ on the 
output signals as shown in Fig.~\ref{figSI6}.
The introduction of statistical noise leads to a
gradual degradation of the computational capacities.
However, the degradation is not abrupt,
which means that QRC would be able to function in a practical situation.
In the small $\tau$ region,
the STM capacity is sensitive to the statistical observational noise. 
This is because in such a region, 
the dynamic range of the observed signals becomes narrow.
For example, when $\tau \Delta = 0.5$ and $\tau \Delta =4$, the dynamic ranges are $\sim 0.01$
and $\sim 0.5$, respectively.
While, in the ideal case, the performances of the 5-qubit QRs 
are comparable to the ESNs of 100 nodes,
their performances
under the statistical observational noise 
of the order of $10^{-3}$
against the dynamic ranges still comparable to the ESNs 
of 50 nodes without any noise.
Moreover, as we saw in the demonstration of the chaotic time series prediction,
we even introduced statistical noise to the observed signals 
with the aim of stabilizing the learning process.
This implies that in some situation we can positively
exploit the natural observational noise in our framework.

These tolerances against imperfections 
indicate that the proposed QRC
framework soundly functions in realistic 
experimental setups as physical reservoir computing. 

\section{Discussion}
The QRC approach enables us to exploit any kind of quantum systems, including quantum simulators and quantum annealing machines, provided their dynamics are sufficiently rich and complex to allow them to be employed for information processing.
In comparison to the standard approach for universal quantum computation,
QRC does not require any sophisticatedly synthesized quantum gate,
but natural dynamics of quantum systems is enough.
Therefore QRC exhibits high feasibility in spite that 
its applications are broad for temporal learning tasks.

The conventional software 
approach for recurrent neural networks
takes a time,
which depends on the size of the network, 
to update the network states.
In contrast, in the case of QRC, 
the time evolution is governed by natural physical dynamics
in a physically parallelized way.
For example, liquid and solid state NMR systems with nuclear and electron spin ensembles~\cite{NMRQC1,NMRQC2} are favourable for implementing QRC. 
These systems enable us to obtain the output signals in real time via the radio-frequency coil by virtue of its huge number of ensembles.
Note that we have employed the simplest model, and that no optimisation of the QRs has been done yet.
More study is necessary to optimise the QRs with respect to a Hamiltonian, network topology, the way of injecting the input sequences, and the readout observables.

Notwithstanding its experimental feasibility, controllability, and robustness against decoherence, the QRC framework would be also useful to analyse complex real-time quantum dynamics from an operational perspective.
The computational capabilities provide operational measures for quantum integrable and chaotic dynamics.
Apparently, the STM is closely related to time correlation in many-body quantum physics and the thermalisation of closed quantum systems.
Moreover, the chaotic behaviour of quantum systems has been investigated in an attempt to understand the fast scrambling nature of black holes~\cite{Scrambler1,Scrambler3}.
It would be intriguing to measure the computational capabilities of such black hole models.
We believe that QRC for universal real-time quantum computing, which bridges quantum information science, machine learning, quantum many-body physics, and high-energy physics coherently, provides an alternative paradigm to quantum digital computing.

\section{Acknowledgements}
K.N. and K.F. are supported by JST PRESTO program. 
K.N. is supported by KAKENHI No. 15K16076 and No. 26880010.
K.F. is supported by KAKENHI No. 16H02211.



\appendix

\section{Experimental Settings and Extended Analyses}
This section describes detailed settings for the task experiments mentioned in the main text and provides extended analyses.
We have maintained the notation for symbols used in the main text.

\subsection{The timer task}
\label{Ap:timer}
The timer task is one of the simplest yet most important benchmark tasks to evaluate the memory capacity of a system (see, e.g., Ref. \cite{Buonomano1}).
As explained in the main text, our goal for the first demonstration of QRC was to emulate the function of a timer (Fig.\ref{fig2} (a) in the main text).
The I/O relation for a timer can be expressed as follows:
\begin{align*}
s_{k} &= \left \{
\begin{array}{l}
1 \ \ (k \geq k') \\
0 \ \ ({\rm otherwise})
\end{array}
\right. \\
y_{k} &= \left \{
\begin{array}{l}
1 \ \ (k=k'+\tau_{\rm timer}) \\
0 \ \ ({\rm otherwise}),
\end{array}
\right.
\end{align*}
where $k'$ is a timestep for launching the cue to the system, and $\tau_{\rm timer}$ is a delay for the timer.
Our aim was to emulate this timer by exploiting the QR dynamics generated by the input projected to the first qubit in the QR system.

A single experimental trial of the task consists of 800 timesteps, where the first 400 timesteps are discarded as initial transients.
At timestep 500, the input is switched from 0 to 1 (i.e. $k' = 500$), and the system continues to run for another 300 timesteps.
For the training procedure, using a 6-qubit QR system with $\tau \Delta = 1$, we iterated this process over five trials, starting from different initial conditions, and collected the corresponding QR time series for each timestep from timestep 400 to timestep 800 as training data. 
We optimised the linear readout weights using these collected QR time series with a linear regression to emulate the target output for the given delay $\tau_{\rm timer}$ and the setting of the number of virtual nodes $V$ in QR systems.
We evaluated the performance of the system with the optimised weights by running five additional trials (evaluation trials) and compared the system outputs to the target outputs in the time region from timestep 400 to timestep 800.

Here, we aim to analyse the performance of the timer task further.
We prepared 10 different 6-qubit QR systems, whose coupling strengths are assigned differently, and for each setting of $(\tau_{\rm timer}, V)$, we iterated the experimental trials as explained above over these 10 different systems.
To effectively evaluate the system's performance against the target outputs $\bar{y}_{k}$, given the setting of $\tau_{\rm timer}$, we defined a measure $C(\tau_{\rm timer})$, which is expressed as
\begin{equation*}
C(\tau_{\rm timer}) = \frac{{\rm cov}^{2}(y_{k}, \bar{y}_{k})}{\sigma^{2}(y_{k})\sigma^{2}(\bar{y}_{k})},
\end{equation*}
where ${\rm cov}(x, y)$ and $\sigma(x)$ express the covariance between $x$ and $y$ and the standard deviation of $x$, respectively.
In short, this measure evaluates the association between two time series, and takes a value from 0 to 1. 
If the value is 1, it means that the system outputs and the target outputs completely overlap, which implies that the learning was perfect.
At the other extreme, if the value is 0, it implies that the learning completely failed.
Evaluation trials were used to actually calculate this measure.
Now, we further define a measure, capacity $C$, which is expressed as a simple summation of $C(\tau_{\rm timer})$ over the entire delay,
\begin{equation*}
C = \sum_{\tau_{\rm timer}=0}^{\tau_{\rm timer}^{\rm max}} C(\tau_{\rm timer}),
\end{equation*}
where $\tau_{\rm timer}^{\rm max}$ is set to 300 in our experiments.

\begin{figure*}
\centering
\includegraphics[width=160mm,clip]{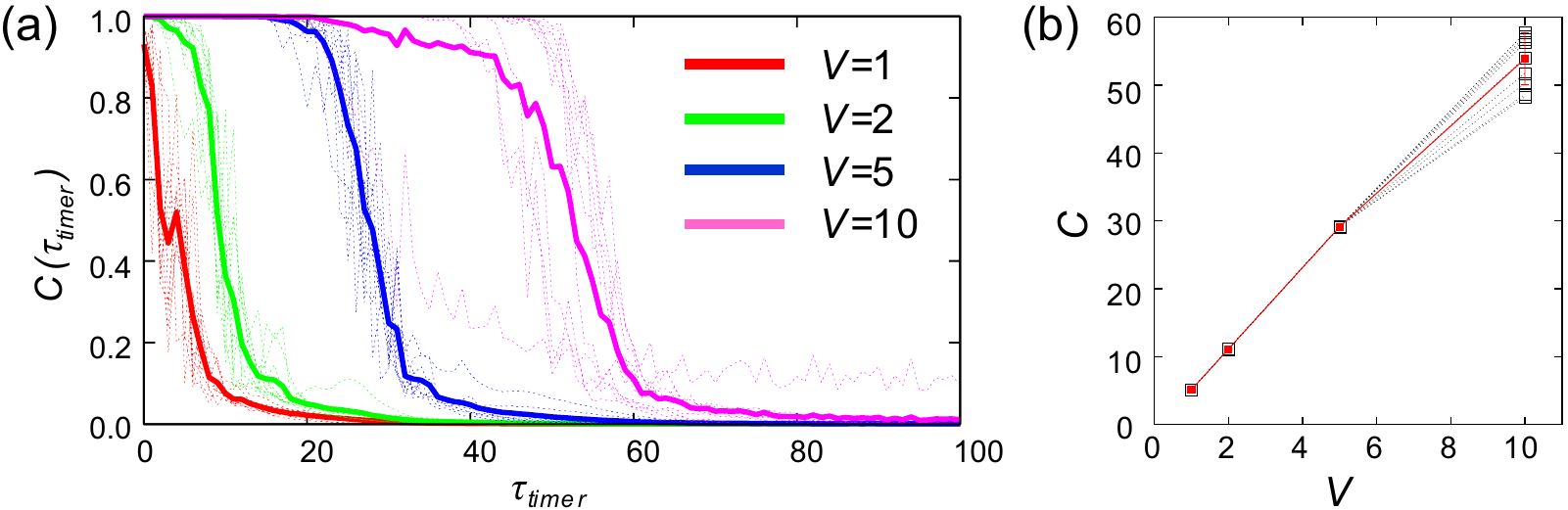}
\caption{Analyses of the memory capacity based on $C(\tau_{\rm timer})$ and $C$ for a 6-qubit QR system.
(a) Plot showing $C(\tau_{\rm timer})$ according to $\tau_{\rm timer}$ for $V=1, 2, 5,$ and $10$. For each setting of $V$, 10 different 6-qubit QR systems are prepared and for each system, the average of $C(\tau_{\rm timer})$ is calculated using five trials according to $\tau_{\rm timer}$ (dashed lines: the cases for 10 different QR systems are overlaid), and the averaged values for these averaged $C(\tau_{\rm timer})$ over 10 different QR systems are overlaid as solid lines.
(b) Plot of $C$ as a function of $V$. Similar to (a), the averaged $C$ for each different QR system for each number of $V$ is overlaid as a dashed line and the averaged values for these averaged $C$ over 10 different QR systems are overlaid as solid lines. The error bars represent the standard deviations.}
\label{figSI1}
\end{figure*}

By using these two measures, $C(\tau_{\rm timer})$ and $C$, we evaluated the performance of the timer tasks of 6-qubit QR systems.
Figure \ref{figSI1} plots the results.
Figure \ref{figSI1} (a) clearly indicates that larger values of $V$ can perform the timer task reliably for a longer delay, which shows a characteristic curve for each setting of $V$.
This point is also confirmed by checking the plot of $C$ according to the value of $V$, where $C$ increases almost linearly with an increase in $V$ (see Fig.\ref{figSI1} (b)).
These results are consistent with the result demonstrated in the main text.

\subsection{The NARMA task}
\label{ap:NARMA}
As explained in the main text, the emulation of NARMA systems is a challenge for machine learning systems in general because it requires nonlinearity and memory \cite{Capacity}.
Thus, the emulation task has been a benchmark for evaluating the amount of nonlinearity and memory to be exploited in the system \cite{benchmark,Reservoir,Laser0,long_short,Helmut0,Kohei0,Kohei1,Kohei3}.
These tasks appear as the second demonstration of QRC in the main text (Fig.\ref{fig2} (b) in the main text).
Here, we explain the experimental procedures in detail and present extended analyses for these tasks.

We used a superimposed sine wave for the input to the NARMA systems, which is expressed as follows:
\begin{equation*}
s_{k} = 0.1 \left( \sin(\frac{2 \pi \alpha k}{T})  \sin(\frac{2 \pi \beta k}{T})  \sin(\frac{2 \pi \gamma k}{T})  + 1   \right),
\end{equation*}
where $(\alpha, \beta, \gamma) = (2.11, 3.73, 4.11)$ and $T=100$.
Note that $s_{k}$ is in the range $[0, 0.2]$ with the aim of stabilizing the behaviour of the NARMA systems (to prevent divergence).
Similar types of input sequences for NARMA systems can be found in Ref. \cite{Helmut0,Kohei0,Kohei1,Kohei3}.
The input range is rescaled to $[0, 1]$ when projected to the first qubit of the QR system. 
The experimental trial consists of 5000 timesteps, where the first 1000 timesteps are used for the washout, the following 3000 timesteps are used for the training phase, and the final 1000 timesteps are used for the evaluation phase.
Note that when the input is a superimposed sine wave, we should be careful to prevent the same input and target output time series in the training phase from appearing again in the evaluation phase, because this would not enable us to characterise the generalisation capability of the system effectively.
Our setting of the length of the training and evaluation phases is confirmed to be safe on this point. 
By collecting the QR time series and the corresponding target outputs for each task in the training phase, we train the linear readouts for five outputs, which correspond to the five target NARMA systems, by using the scheme explained in the main text.
The trained linear readouts are used to generate system outputs for the evaluation phase.

The contribution of the QR system is characterised explicitly by comparing the task performance with a simple linear regression (LR) model, $y_{k+1} = w'_{1} \times s_{k} + w'_{0}$, where $w'_{0}$ and $w'_{1}$ are trained using the same time series as in the training phase.
Note that this corresponds to the case in which no QR system is available, and only the raw input remains for LR.
This comparison enables us to conclude that, for any system performance exceeding that of this model, the QR system has contributed to the emulation task \cite{Helmut0,Kohei0,Kohei1,Kohei3,Kohei2}.

We evaluate the performance of the system output in the evaluation phase by calculating the normalised mean squared error (NMSE) with the target output: 
\begin{equation}
NMSE = \frac{\sum_{k=L+1}^{M-L}(\bar{y}_{k+1}-y_{k+1})^{2}}{\sum_{k=L+1}^{M-L}\bar{y}_{k+1}^{2}},
\end{equation}
where $L$ represents the timesteps for the washout and training phase, of which the duration is 4000 timesteps in this experiment, and $M$ is the timesteps for the evaluation phase, which requires 1000 timesteps.
Table I lists the NMSE for each of the experimental conditions.
We can confirm that our 6-qubit QR system outperforms the LR system in any setting of $V$ for each NARMA task, which implies that the QR system has contributed to the task performance.
Furthermore, we can see that by increasing $V$, the performance improves in all the NARMA tasks, which is consistent with the plots presented in Fig.\ref{fig2} (b) in the main text.

\begin{table}[htbp]
\begin{center}
\caption{Performance of the 6-qubit QR systems in terms of NMSE for NARMA tasks using the superimposed sine wave.}
\small
\scalebox{0.8}[0.85]{
\begin{tabular}{|c|c|c|} \hline
\multicolumn{1}{|c|}{Task} & \multicolumn{1}{|c|}{System} & \multicolumn{1}{c|}{Error (NMSE)} \\ \hline
NARMA2 & LR & $1.7 \times 10^{-5}$ \\ \cline{2-3}
 & QR ($V=1$) & $1.0 \times 10^{-5}$ \\ \cline{2-3}
 & QR ($V=2$) & $4.7 \times 10^{-6}$ \\ \cline{2-3}
 & QR ($V=5$) & $1.7 \times 10^{-7}$ \\ \cline{2-3}
 & QR ($V=10$) & $4.9 \times 10^{-8}$ \\ \cline{2-3} \hline
NARMA5 & LR & $3.0 \times 10^{-3}$ \\ \cline{2-3}
 & QR ($V=1$) & $4.6 \times 10^{-4}$ \\ \cline{2-3}
 & QR ($V=2$) & $7.1 \times 10^{-5}$ \\ \cline{2-3}
 & QR ($V=5$) & $2.8 \times 10^{-5}$ \\ \cline{2-3}
 & QR ($V=10$) & $7.6 \times 10^{-6}$ \\ \cline{2-3} \hline
NARMA10 & LR & $2.6 \times 10^{-3}$ \\ \cline{2-3}
 & QR ($V=1$) & $2.0 \times 10^{-4}$ \\ \cline{2-3}
 & QR ($V=2$) & $9.2 \times 10^{-5}$ \\ \cline{2-3}
 & QR ($V=5$) & $3.0 \times 10^{-5}$ \\ \cline{2-3}
 & QR ($V=10$) & $1.3 \times 10^{-5}$ \\ \cline{2-3} \hline
NARMA15 & LR & $2.7 \times 10^{-3}$ \\ \cline{2-3}
 & QR ($V=1$) & $6.7 \times 10^{-4}$ \\ \cline{2-3}
 & QR ($V=2$) & $3.1 \times 10^{-4}$ \\ \cline{2-3}
 & QR ($V=5$) & $1.2 \times 10^{-4}$ \\ \cline{2-3}
 & QR ($V=10$) & $4.0 \times 10^{-5}$ \\ \cline{2-3} \hline
NARMA20 & LR & $2.3 \times 10^{-3}$ \\ \cline{2-3}
 & QR ($V=1$) & $1.2 \times 10^{-3}$ \\ \cline{2-3}
 & QR ($V=2$) & $2.6 \times 10^{-4}$ \\ \cline{2-3}
 & QR ($V=5$) & $1.3 \times 10^{-4}$ \\ \cline{2-3}
 & QR ($V=10$) & $3.8 \times 10^{-5}$ \\ \hline
\end{tabular}
}
\end{center}
\label{table1}
\end{table}

Here we aim to further analyse the information processing capacity of our QR system based on the NARMA tasks.
We adopt the same task settings as for the previous case except for the input settings.
The input stream is generated by using white noise with a range of $[0, 0.2]$ for the same reason as in the previous experiment, rather than using a superimposed sine wave.
This choice of input stream is commonly used \cite{Reservoir,Laser0,benchmark} and is determined not to add additional temporal coherence originating from external input to the system, and to evaluate the pure computational power only contributed by the QR systems.
As this input setting perceivably complicates the performance evaluation, we quantified the task performance in terms of NMSE.
For each NARMA task, we tested the relevance of $\tau$ and $V$ in terms of the task performance and varied them for $\tau \Delta = 1, 2, 4, 8, 16, 32, 64,$ and $128$, and $V=1, 2, 5, 10, 25,$ and $50$, respectively.
Using a 5-qubit QR system, 20 samples of the QRs were randomly generated and for each $(\tau \Delta, V)$ setting, the average values of NMSEs were obtained.

\begin{figure}
\centering
\includegraphics[width=80mm,clip]{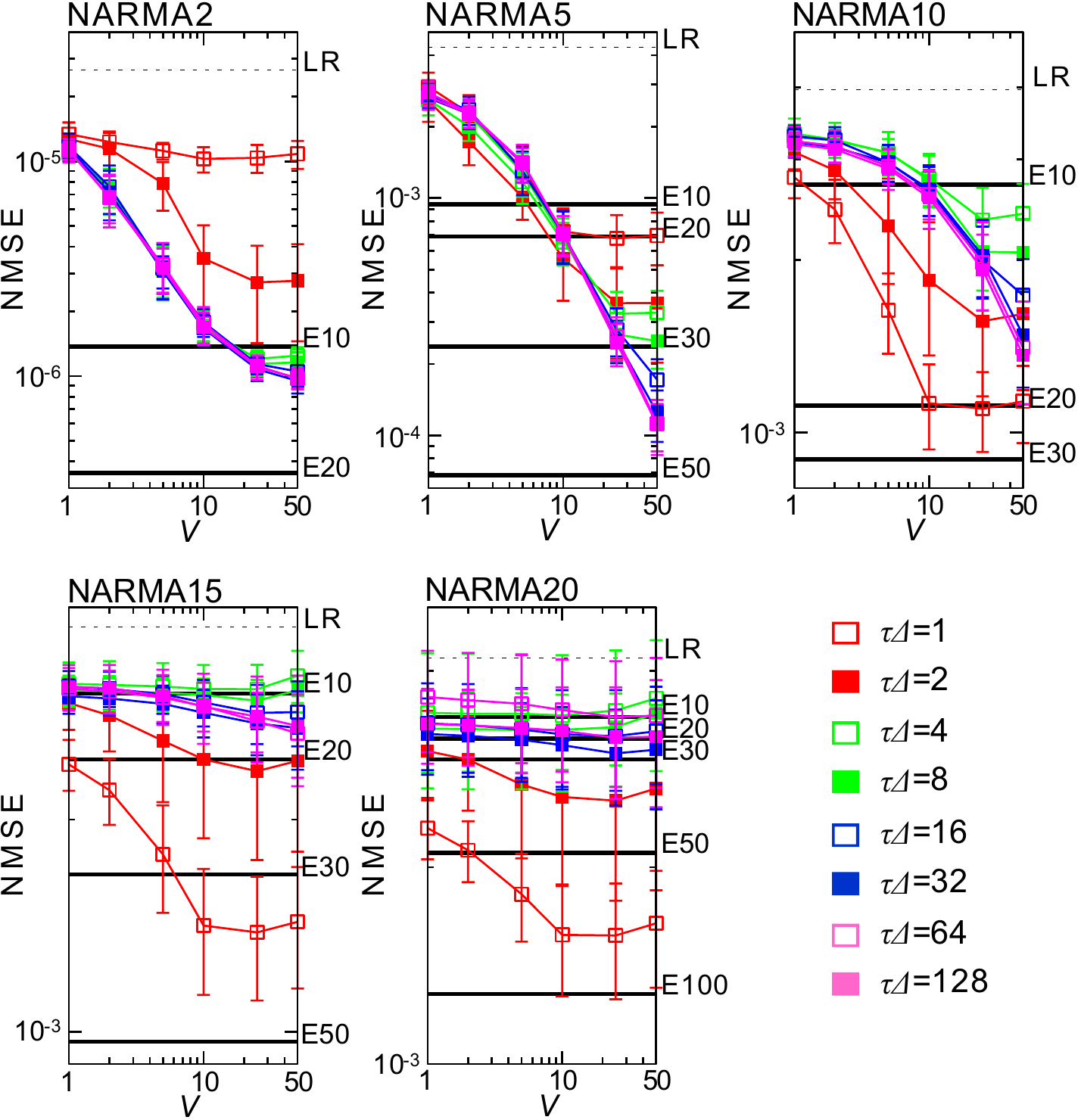}
\caption{Performance of 5-qubit QR systems with several $\tau \Delta$ settings for the NARMA tasks with a random input stream. 
For each plot, the vertical and horizontal axes show the NMSE and the number of virtual nodes $V$, respectively, and both axes are in logarithmic scales.
The error bars show the standard deviations evaluated on 20 samples of the QRs with respect to the random couplings.
For comparisons, the dashed line shows the performance of an LR system in terms of NMSE, and the solid lines show the performance of an ESN with $N$ nodes (e.g. ``E10'' represents the performance of ESN with 10 nodes).
See text for details on analyses and experimental procedures.}
\label{figSI2}
\end{figure}

The performance of the QR systems was characterised by again using the previously mentioned LR system for comparison. 
Furthermore, we used a conventional echo state network (ESN) as a candidate for the standard machine learning system and used it to compare our task performance.
The basic settings of the ESN are described in Appendix~\ref{Ap:ESN}.
To allow for fair comparisons, 100 samples of ESN with $N$ nodes were generated by assigning the same NARMA tasks with the same experimental settings explained above, and the average values of NMSEs for the ESN were obtained (by varying the spectral radius of the ESN internal weight matrix from 0.05 to 1.95 in increments of 0.1; the case of the smallest NMSE, which provided the best performance, was used for comparison).
The number of nodes $N$ was varied for $N=10, 20, 30, 50,$ and $100$ for each NARMA task.

Figure \ref{figSI2} depicts the results for the performance of the 5-qubit QR system for the NARMA tasks with a random input stream.
First, we can see that for all the NARMA tasks, our 5-qubit QR system outperformed the LR system, which means that the QR system actually contributed to the task performance.
In general, we can confirm that the increase in the number of the virtual nodes $V$ leads to an improved performance.
The effect of the number of $\tau \Delta$ on the task performance depends on the type of task.
For the NARMA2 and NARMA5 tasks, an increase in the number of $\tau \Delta$ tends to improve the performance.
In particular, for the NARMA5 task, when $V=50$ and $\tau \Delta=128$, the 5-qubit QR system performed at a level in between the performance of the ESN with $N=30$ and $N=50$ (see Fig. \ref{figSI2}).
For NARMA10, NARMA15, and NARMA20, this tendency does not hold in general and the use of a small number of $\tau \Delta$ was found to improve the performance.
In particular, for the NARMA20 task, when $\tau \Delta=1$ and $V>5$, the 5-qubit QR system performed at a level in between the performance of the ESN with $N=50$ and $N=100$ (see Fig. \ref{figSI2}).
Interestingly, this implies that the 5-qubit QR system can deliver a performance similar to that of an ESN with fewer computational nodes (e.g. when $\tau \Delta=1$ and $V=5$, the 5-qubit QR system has 25 computational nodes and the performance exceeds that of an ESN with 50 nodes).
These outcomes of the task performance are induced by the balancing of memory and nonlinearity, which can be exploited by the system and which is required to perform the task; this is closely related to the results mentioned in the main text.
Further analyses will be included in our future work.

\subsection{The Mackey-Glass prediction task}
\label{Ap:MG}
Chaotic attractor learning is a popular test for learning dynamical systems \cite{Jaeger0,Jaeger0SI}. 
One of the well-known systems used for the target of learning is the Mackey-Glass (MG) delay differential equation
\begin{equation*}
\dot{y} (t) = \frac{\alpha y (t-\tau_{\rm MG})}{1+y (t-\tau_{\rm MG})^{\beta}} - \gamma y (t),
\end{equation*}
where the parameters are set to $\alpha = 0.2$, $\beta = 10$, and $\gamma = 0.1$. 
The system has a chaotic attractor if $\tau_{\rm MG} > 16.8$.
In the majority of studies, $\tau_{\rm MG} = 17$ is used, which yields a chaotic attractor.
In our experiments, we also used this parameter setting of $\tau_{\rm MG}=17$.
Additionally, we used the case of $\tau_{\rm MG}=16$ for comparison, as this setting does not exhibit chaos.

The discrete time version of the MG system is often used to prepare the training sequences \cite{Jaeger0} through
\begin{equation*}
y_{k+1} = y_{k} + \sigma \left( \frac{0.2 y_{k-\frac{\tau_{\rm MG}}{\sigma}}}{1+y_{k-\frac{\tau_{\rm MG}}{\sigma}}^{10}} - 0.1 y_{k} \right),
\end{equation*}
with a step size of $\sigma = 1/10$ and then sub-sampled by $10$. 
One step from $k$ to $k + 1$ in the resulting sequences corresponds to a unit time interval $[t, t + 1]$ of the original continuous system.
In our numerical experiments, the target time series is linearly scaled to $[0, 1]$ and used in the actual experiments.

For each setting of $\tau_{\rm MG}$, we generated the above system for a while as a washout and then a length of 12000 timesteps (already sub-sampled) was collected for the experiment.
We used 10000 timesteps for the training phase and the remaining 2000 timesteps for the evaluation phase. 
The task was to train the QR system by using these training sequences, which after training should re-generate the corresponding chaotic or non-chaotic attractors.

Because this task requires feedback to the system, the training procedure is different from the previous cases.
During the training phase, we clamped the feedback from the system output, and provided the target outputs as inputs, which means we set $s_{k+1} = \bar{y}_{k}$. 
Thus, the training phase was carried out with an open loop, such that the system was forced into the desired operative state by the target signals (this approach is typically referred to as {\it teacher forcing}).
The robustness of the learning was improved by adding a slight amount of noise in the range of $[-\sigma, \sigma]$ in the training phase. 
When the evaluation phase started, we switched the inputs to the system output generated by the trained readout weights (this phase is expressed as the autonomous phase in Fig. \ref{fig2} (c) in the main text) and checked whether the system was able to embed the corresponding MG system.

Table II summarises the experimental conditions and the prediction errors for the QR system used in the main text.
We calculated the errors in NMSE by using the entire time series in the evaluation phase.

\begin{table}[htbp]
\begin{center}
\caption{Experimental settings and prediction errors (NMSE) for the Mackey-Glass prediction tasks in the main text.}
\small
\scalebox{0.8}[0.85]{
\begin{tabular}{|c|c|c|c|c|c|} \hline
\multicolumn{1}{|c|}{$\tau_{\rm MG}$} & \multicolumn{1}{|c|}{Case} & \multicolumn{1}{|c|}{Qubit} & \multicolumn{1}{|c|}{$\tau \Delta$} & \multicolumn{1}{c|}{Noise strength ($\sigma$)} & \multicolumn{1}{c|}{Error (NMSE)} \\ \hline
16 & 1 & 6 & 3 & $1.0 \times 10^{-4}$ & $4.7 \times 10^{-3}$ \\ \cline{2-6}
 & 2 & 7 & 2 & $1.0 \times 10^{-4}$ & $3.9 \times 10^{-3}$ \\ \hline
17 & 1 & 6 & 3 & $1.0 \times 10^{-4}$ & $1.6 \times 10^{-1}$ \\ \cline{2-6}
 & 2 & 7 & 3 & $1.0 \times 10^{-4}$ & $2.5 \times 10^{-2}$ \\ \cline{2-6}
 & 3 & 7 & 4 & $1.0 \times 10^{-5}$ & $4.9 \times 10^{-2}$ \\ \cline{2-6}
 & 4 & 7 & 2 & $1.0 \times 10^{-5}$ & $1.7 \times 10^{-2}$ \\ \hline
\end{tabular}
}
\end{center}
\end{table}

\begin{figure}
\centering
\includegraphics[width=80mm,clip]{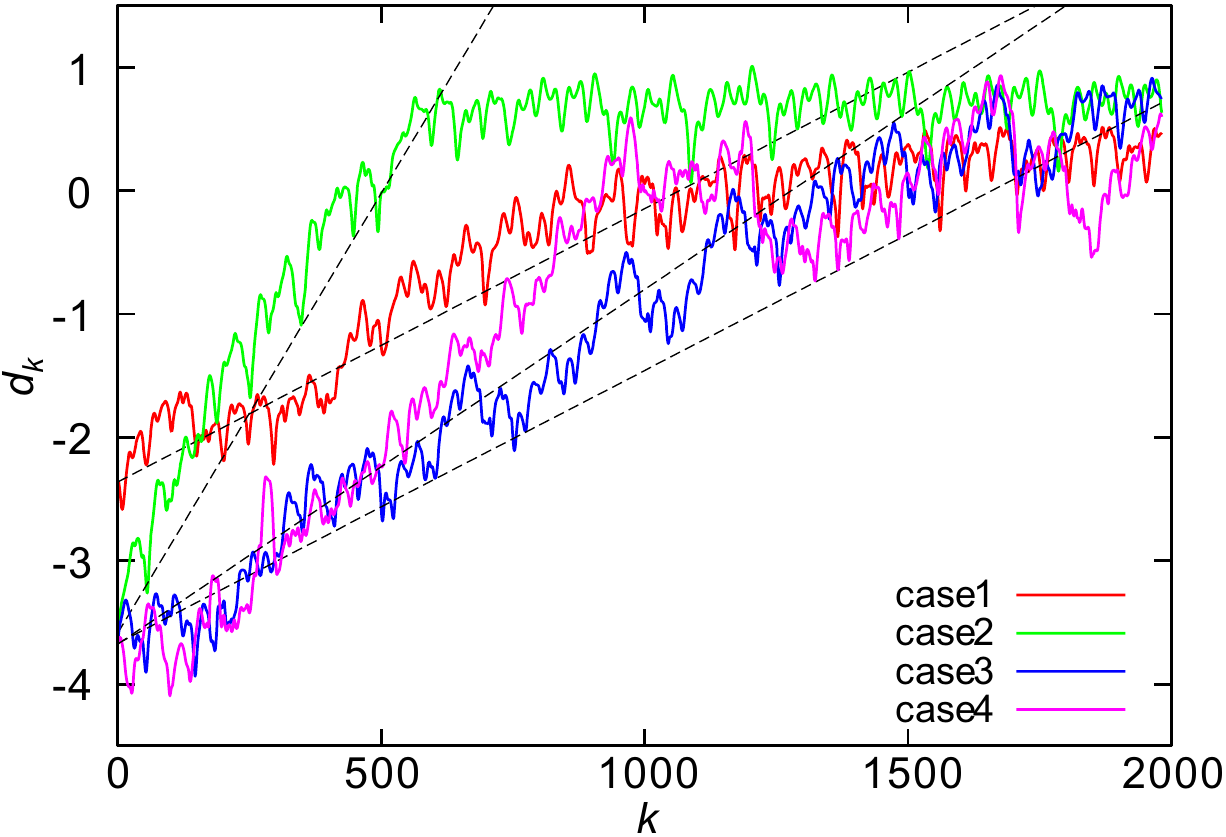}
\caption{Plot showing the time series of $d_{k}$.
The results for case 1 $\sim$ 4 are overlaid.
Note that the vertical axis is in a logarithmic scale.
The estimated largest Lyapunov exponent $\lambda$ is 0.0022, 0.0071, 0.0022, and 0.0029 for case 1, 2, 3, and 4, respectively.}
\label{figSI3}
\end{figure}

We tested whether the trained network indeed generates a chaotic time series by empirically estimating the largest Lyapunov exponent of the network-generated output signal by using a procedure similar to that introduced in Ref. \cite{Jaeger0SI}. 
For the trained network, we analysed the previous four cases (case 1 $\sim$ 4) in $\tau_{\rm MG} = 17$ setting. 
When the network was switched from the teacher forcing condition to the closed-loop mode at timestep 10000, 
the reservoir signals were perturbed by a uniform noise vector, and the network was left running freely, on this occasion starting from the perturbed state for the entire 2000 steps of the evaluation phase, and the resulting output sequence was recorded. 
The exponential divergence rate $\lambda$ between this perturbed sequence $y'_{k}$ and the original sequence $y_{k}$ was estimated by computing
\begin{align*}
d_{k} &= \| [y_{10001+k} ... y_{10017+k}]-[y'_{10001+k} ... y'_{10017+k}] \|, \\
\lambda &= \frac{\log (d_{500}) - \log (d_{0})}{500},
\end{align*}
where the subsequent 17 timesteps that are used for the computation of $d_{k}$ were chosen because they correspond to approximately one full ``loop'' of the attractor. 
Figure \ref{figSI3} plots the behaviour of $d_{k}$ for four cases.
We can see that all four cases have a positive $\lambda$ value, which implies that their output sequences are chaotic.

\subsection{Echo state network settings for comparisons}
\label{Ap:ESN}
\begin{figure}
\centering
\includegraphics[width=80mm,clip]{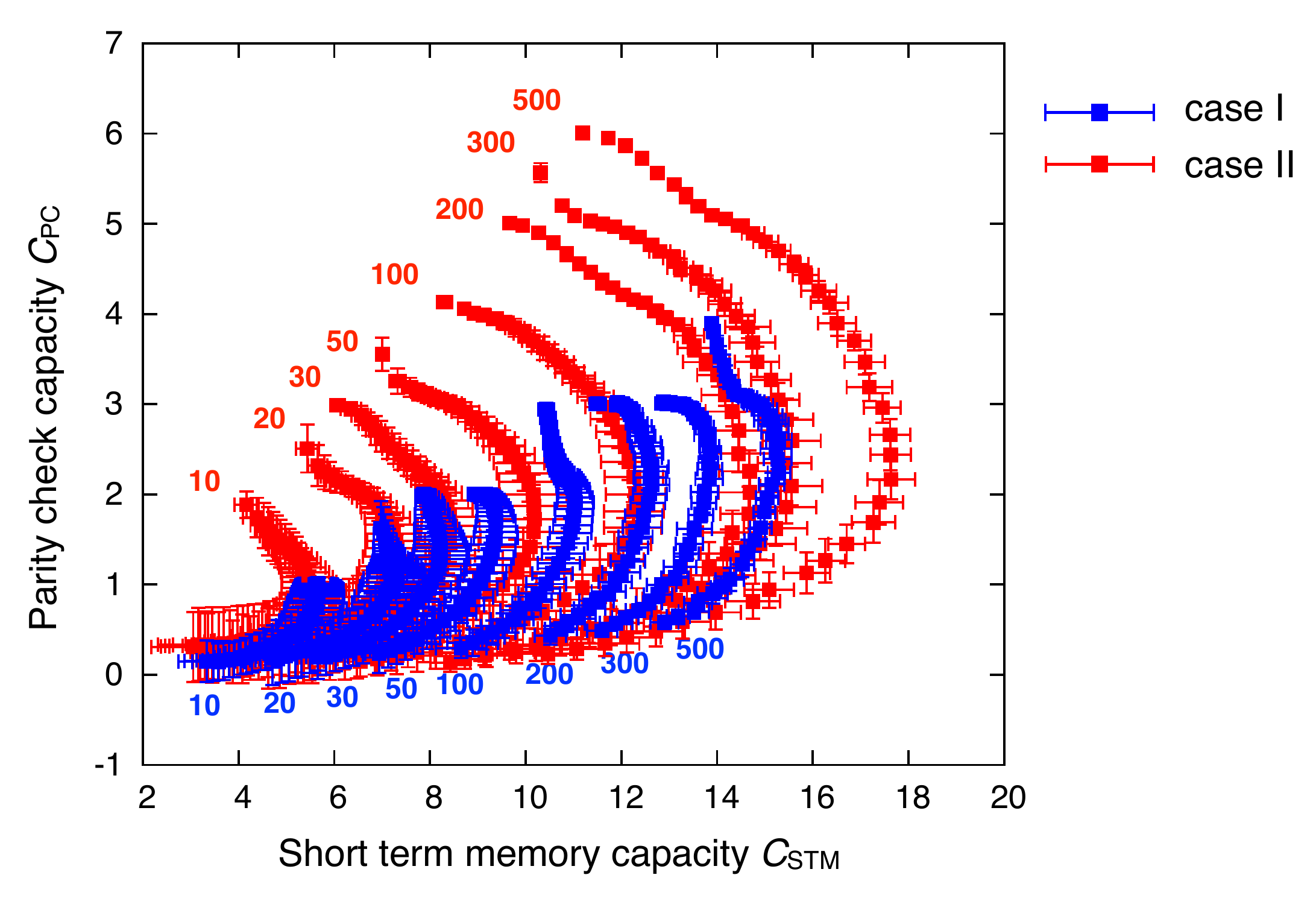}
\caption{Capacities $(C_{\rm STM},C_{\rm PC})$ of the ESNs with two types of input sequence.
Case I and II represent input type with $\{ -1,1 \}$ (coloured blue) and with $\{ 0,1 \}$ (coloured red), respectively. 
Plots show the results for the spectral radius from 0.05 to 2.00.
The error bars show the standard deviations 
evaluated on 100 samples of the randomly generated ESNs.}
\label{figSIXX}
\end{figure}
We further characterised the computational power of our QR system by comparing its task performance with that of a conventional echo state network (ESN) \cite{Jaeger0,Jaeger0SI,Jaeger1,Jaeger3} (the comparisons of ESN performance with that of our systems appear in Sec.~\ref{sec:character} and also in the analyses of the NARMA tasks demonstrated above).
The ESN is a type of recurrent neural network, which has $N$ internal network units, input units, and output units.
Activation of the $i$th internal unit at timestep $k$ is $x^{i}_{k}$ ($i=1, ..., N$). 
We used the same I/O setting for the ESN as with our system for each task concerned to enable us to efficiently and directly compare the performance.
The number of trials, the lengths of the washout, training, and evaluation phases, and the evaluation procedures are also kept the same.
The connection weights for the $N \times N$ internal network connecting the $i$th unit with the $j$th unit are denoted as $w_{ij}$, and the input weights proceeding from the input unit into the $i$th internal unit are denoted as $w_{\rm in}^{i}$.
The readout weights $w^{i}_{\rm out}$ proceed from $N$ internal units and one bias to the output unit (where $x^{0}_{k}=1$ and $w^{0}_{\rm out}$ for a bias term).
The readout weights $w^{i}_{\rm out}$ are trained using the procedure explained for each task; the internal weights $w_{ij}$ and the input weights $w_{\rm in}^{i}$ are randomly assigned from the range $[-1.0, 1.0]$ and fixed beforehand.
The activation of the internal units and the output unit are updated as
\begin{align*}
x^{i}_{k} &= f (\sum_{j=1}^{N} w_{ij}x^{j}_{k-1} + w_{\rm in}^{i}s_{k} ), \\
y_{k} &= \sum_{i=0}^{N} w^{i}_{\rm out}x^{i}_{k},
\end{align*}
where $f$ is a $\tanh$ function.
It is reported that the computational power of ESN can be well characterised by the spectral radius of the internal connection weight matrix \cite{Reservoir,Jaeger0,Jaeger0SI,Jaeger1,Jaeger3}.
In each comparative experiment, by incrementally varying the spectral radius, we observed the ESN performance.
Detailed experimental conditions are given for each of these comparisons.

Here we present the ESN settings for the comparisons with QR systems appeared in Sec.~\ref{sec:character}.
The experimental settings including the length of training and evaluation phases are kept the same with the QR system for the fair comparison.
For the random binary input sequence, we adopted two cases.
In the first case (Case I), we changed the actual input value to ``$-$1'' only if $s_{k} = 0$.
For the second case (Case II), we directly projected the $\{ 0,1 \}$-binary state input $s_{k}$ to the internal network units.
In the ESN, if $s_{k} = 0$, the internal units receive no external input, and therefore, are expected to introduce an asymmetry into the network performance.   
We tested these two cases (Case I and Case II) in Fig.~\ref{figSIXX}.
As can be seen from the plot, both cases show different modalities of performances in terms of $C_{\rm PC}$ and $C_{\rm STM}$, which are due to the asymmetry introduced by the input settings.
We have presented the results for Case I in Sec.~\ref{sec:character} but the same explanations hold for both cases.


\begin{thebibliography}{99}

\bibitem{Feynman}
R.P. Feynman, 
{\it Simulating physics with computers},
Int. J. Theor. Phys. {\bf 21}, 467 (1982).

\bibitem{NielsenChuang}
M.A. Nielsen and I. L. Chuang,
{\it Quantum computation and quantum information}, 
(Cambridge university press 2010).

\bibitem{FujiiText}
K. Fujii,
{\it Quantum Computation with Topological Codes
-From Qubit to Topological Fault-Tolerance-},
SpringerBriefs in Mathematical Physics
(Springer-Verlag 2015).


\bibitem{ShorFact}
P. W. Shor, 
{\it Algorithms for quantum computation: Discrete logarithms and factoring},
In Proceedings of the 35th Annual Symposium on Foundations of Computer Science, 
124 (1994).


\bibitem{UCSB1}
R. Barends {\it et al.,}, 
{\it Superconducting quantum circuits at the surface code threshold for fault tolerance}, 
Nature {\bf 508}, 500 (2014).

\bibitem{UCSB2}
J. Kelly {\it et al.,} 
{\it State preservation by repetitive error detection 
in a superconducting quantum circuit}, 
Nature {\bf 519}, 66 (2015). 

\bibitem{QuantumSimulator1}
J.I. Cirac and P. Zoller, 
{\it Goals and opportunities in quantum simulation},
Nat. Phys. {\bf 8}, 264 (2012).

\bibitem{QuantumSimulator2}
I. Bloch, J. Dalibard, and S. Nascimb{\`e}ne,
{\it Quantum simulations with ultracold quantum gases},
Nat. Phys. {\bf 8}, 267 (2012).

\bibitem{QuantumSimulator3}
I. M. Georgescu, S. Ashhab, and F. Nori,
{\it Quantum simulation},
Rev. of Mod. Phys. {\bf 86}, 153 (2014).

\bibitem{Nishimori}
T. Kadowaki and H. Nishimori, 
{\it Quantum annealing in the transverse Ising model},
Phys. Rev. E {\bf 58}, 5355 (1998).

\bibitem{Farhi}
E. Farhi {\it et al.,},
{\it A quantum adiabatic evolution algorithm applied to random instances of an NP-complete problem},
Science {\bf 292}, 472 (2001).

\bibitem{Dwave}
T. F. Ronnow, {\it et al.}, 
{\it Defining and detecting quantum speedup},
Science {\bf 345}, 420 (2014).

\bibitem{Dwave2}
S. Boixo {\it et al.,} 
{\it Evidence for quantum annealing with more than one hundred qubits},
Nat. Phys. {\bf 10}, 218 (2014).

\bibitem{FujiiMorimae}
T. Morimae, K. Fujii, and J. F. Fitzsimons,
{\it Hardness of classically simulating the one-clean-qubit model},
Phys. Rev. Lett. {\bf 112}, 130502 (2014).

\bibitem{Fujiietal}
K. Fujii {\it et al.,}
{\it Power of Quantum Computation with Few Clean Qubits},
Proceedings of 43rd International Colloquium on Automata, Languages, and Programming (ICALP 2016), pp.13:1-13:14.

\bibitem{FujiiTamate}
K. Fujii and S. Tamate,
{\it Computational quantum-classical boundary of complex and noisy quantum systems},
Sci. Rep. {\bf 6}, 25598 (2016).


\bibitem{IBM_Neuro}
P.A. Merolla, {\it et al.},
{\it A million spiking-neuron integrated circuit with a scalable communication network and interface},
Science {\bf 345}, 668 (2014).

\bibitem{Jaeger0} 
H. Jaeger and H. Haas, 
{\it Harnessing nonlinearity: predicting chaotic systems and saving energy in wireless communication},
Science {\bf 304}, 78 (2004).

\bibitem{Maass0} 
W. Maass, T. Natschl\"{a}ger, and H. Markram, 
{\it Real-time computing without stable states: a new framework for neural computation based on perturbations},
Neural Comput. {\bf 14}, 2531 (2002).

\bibitem{Reservoir} 
D. Verstraeten, B. Schrauwen, M. D'Haene, and D. Stroobandt, 
{\it An experimental unification of reservoir computing methods},
Neural Netw. {\bf 20}, 391 (2007).


\bibitem{Transient} 
M. Rabinovich, R. Huerta, and G. Laurent,
{\it Transient dynamics for neural processing}
Science {\bf 321}, 48 (2008). 


\bibitem{Capacity} 
J. Dambre {\it et al.,} 
{\it Information processing capacity of dynamical systems},
Sci. Rep. {\bf 2}, 514 (2012).


\bibitem{Bucket} 
C. Fernando and S. Sojakka, 
{\it Pattern recognition in a bucket}
In Lecture Notes in Computer Science {\bf 2801}, p. 588 (Springer, 2003).

\bibitem{Laser0} 
L. Appeltant {\it et al.,}
{\it Information processing using a single dynamical node as complex system.}
Nat. Commun. {\bf 2}, 468 (2011).

\bibitem{Laser1} 
D. Woods and T. J. Naughton, 
{\it Photonic neural networks.}, 
Nat. Phys. {\bf 8}, 257 (2012).

\bibitem{Laser1b} 
L. Larger {\it et al.,} 
{\it Photonic information processing beyond Turing: an optoelectronic implementation of reservoir computing},
Optics Express {\bf 20}, 3241 (2012).

\bibitem{Laser2} 
Y. Paquot {\it et al.,}
{\it Optoelectronic Reservoir Computing},
Sci. Rep. {\bf 2}, 287 (2012).

\bibitem{Laser3} 
D. Brunner {\it et al.,} 
{\it Parallel photonic information processing at gigabyte per second data rates using transient states},
Nat. Commun. {\bf 4}, 1364 (2013).

\bibitem{Laser4} 
K. Vandoorne {\it et al.,} 
{\it Experimental demonstration of reservoir computing on a silicon photonics chip}
Nat. Commun. {\bf 5}, 3541 (2014).

\bibitem{Neuromorphic0} 
A. Z. Stieg {\it et al.,} 
{\it Emergent criticality in complex turing B-type atomic switch networks},
Adv. Mater. {\bf 24}, 286 (2012).


\bibitem{Helmut0} 
H. Hauser {\it et al.,}
{\it Towards a theoretical foundation for morphological computation with compliant bodies}
Biol. Cybern. {\bf 105}, 355 (2011).

\bibitem{Kohei0}
K. Nakajima {\it et al.}, 
{\it Computing with a Muscular-Hydrostat System},
Proceedings of 2013 IEEE International Conference 
on Robotics and Automation (ICRA), 1496 (2013).


\bibitem{Kohei1} 
K. Nakajima {\it et al.}, 
{\it A soft body as a reservoir: case studies in a dynamic model of octopus-inspired soft robotic arm}
Front. Comput. Neurosci. {\bf 7}, 1 (2013).

\bibitem{Kohei2} 
K. Nakajima {\it et al.,} 
{\it Exploiting short-term memory in soft body dynamics as a computational resource},
J. R. Soc. Interface {\bf 11}, 20140437 (2014).

\bibitem{Kohei3} 
K. Nakajima {\it et al.,} 
{\it Information processing via physical soft body},
Sci. Rep. {\bf 5}, 10487 (2015).

\bibitem{Ken1} 
K. Caluwaerts {\it et al.,} 
{\it Design and control of compliant tensegrity robots through simulations and hardware validation},
J. R. Soc. Interface {\bf 11}, 20140520 (2014).

\bibitem{DL} 
Y. LeCun, Y. Bengio, and G. Hinton, 
{\it Deep Learning} 
Nature {\bf 521}, 436 (2015).

\bibitem{Briegel12}
H. J. Briegel, and G. De las Cuevas, 
{\it Projective simulation for artificial intelligence},
Sci. Rep. {\bf 2}, 400 (2012). 


\bibitem{Briegel14}
G. D. Paparo {\it et al.,}
{\it Quantum speedup for active learning agents}, 
Phys. Rev. X {\bf 4}, 031002 (2014).

\bibitem{Lloyed14}
P. Rebentrost, M. Mohseni, and S. Lloyd,  
{\it Quantum support vector machine for big data classification}, 
Phys. Rev. Lett. {\bf 113}, 130503 (2014).

\bibitem{LloyedNatPhys}
S. Lloyd, M. Mohseni, and P. Rebentrost,
{\it Quantum principal component analysis},
Nat. Phys. {\bf 10}, 631 (2014).

\bibitem{QauntumDeep}
N. Wiebe, A. Kapoor, and K. M. Svore,
{\it Quantum Deep Learning},
arXiv:1412.3489.

\bibitem{QLearning}
J. C. Adcock {\it et al.},
{\it Advances in quantum machine learning},
arXiv:1512.02900.

\bibitem{Wigner}
E. P. Wigner, 
{\it On the statistical distribution of the widths and spacings of nuclear resonance levels}, 
Mathematical Proceedings of the Cambridge Philosophical Society. {\bf 47}, (Cambridge University Press, 1951).

\bibitem{RMT}
T. Guhr, A. M{\"u}llerGroeling, \& H. A. Weidenm{\"u}ller, 
{\it Random-matrix theories in quantum physics: common concepts}, 
Phys. Rep. {\bf 299}, 189 (1998).

\bibitem{QWalk2}
A. M. Childs {\it et al.,}
{\it Exponential algorithmic speedup by a quantum walk},
Proceedings of the 35th ACM Symposium on Theory of Computing (ACM, New York, 2003), pp. 59-68.

\bibitem{QWalk1}
A. M. Childs,
{\it Universal Computation by Quantum Walk},
Phys. Rev. Lett. {\bf 102}, 180501 (2009).

\bibitem{NMRQC1}
D. G. Cory {\it et al.,} 
{\it NMR Based Quantum Information Processing: Achievements and Prospects}, 
Fortschr. Phys. {\bf 48}, 875 (2000). 

\bibitem{NMRQC2}
J. A. Jones, 
{\it Quantum Computing with NMR},
Prog. Nucl. Magn. Reson. Spectros. {\bf 59}, 91 (2011).


\bibitem{randomIsing}
B. Georgeot and D. L. Shepelyansky, 
{\it Quantum chaos border for quantum computing}, 
Phys. Rev. E {\bf 62}, 3504 (2000).


\bibitem{Buonomano1} 
R. Laje and D. V. Buonomano, 
{\it Robust timing and motor patterns by taming chaos in recurrent neural networks}, 
Nat. Neurosci. {\bf 16}, 925 (2013). 

\bibitem{long_short} 
S. Hochreiter and J. Schmidhuber, 
{\it Long short-term memory}, 
Neural Comput. {\bf 9}, 1735 (1997).

\bibitem{benchmark} 
A. F. Atiya and A. G. Parlos, 
{\it New results on recurrent network training: Unifying the algorithms and accelerating convergence}, 
IEEE Trans. Neural Netw. {\bf 11}, 697 (2000).

\bibitem{EdgeofChaos} 
N. Bertschinger and T. Natschl\"{a}ger, 
{\it Real-time computation at the edge of chaos in recurrent neural networks}, 
Neural Comput. {\bf 16}, 1413 (2004).

\bibitem{Scrambler1}
P. Hayden and J. Preskill,
{\it Black holes as mirrors: quantum information in random subsystems},
J. High Energy Phys. {\bf 2007}, 120 (2007).

\bibitem{Scrambler3}
P. Hosur {\it et al.,}
{\it Chaos in quantum channels},
J. High Energ. Phys. {\bf 2016}, 4 (2016).

\bibitem{Jaeger1} 
H. Jaeger, 
{\it Tutorial on training recurrent neural networks, covering BPTT, RTRL, EKF and the ``echo state network'' approach},
GMD Report {\bf 159}, German National Research Center for Information Technology (2002).

\bibitem{Jaeger0SI} 
H. Jaeger,
{\it The ``echo state'' approach to analysing and training recurrent neural networks}, 
GMD Report {\bf 148}, German National Research Institute for Computer Science (2001).

\bibitem{Jaeger3} 
H. Jaeger,
{\it Short term memory in echo state networks},
GMD Report {\bf 152}, German National Research Center for Information Technology (2001).
\end{thebibliography}
\end{document}